\documentclass[10pt,conference]{IEEEtran}
\IEEEoverridecommandlockouts

\usepackage[normalem]{ulem}
\usepackage[utf8]{inputenc}
\usepackage{amsmath}  
\usepackage{graphicx}
\usepackage{comment}
\usepackage{multirow}
\usepackage{enumitem}
\usepackage{array}

\usepackage{xcolor}
\usepackage{balance}
\usepackage{mathtools}
\usepackage{xspace}
\usepackage{url}
\usepackage{eucal}
\usepackage{amsfonts}
\usepackage{color}
\usepackage{booktabs}
\usepackage{bbding}
\usepackage{float}

\usepackage{graphicx}
\usepackage{hyperref}
\usepackage{url}
\usepackage{multirow}
\usepackage{comment}
\usepackage{colortbl}
\usepackage{graphicx}
\usepackage{subcaption}
\usepackage[linewidth=0.1pt]{mdframed}
\usepackage[linesnumbered,ruled,lined]{algorithm2e}

\usepackage{graphicx}
\usepackage{hyperref}
\usepackage{url}
\usepackage{multirow}
\usepackage{comment}
\usepackage{colortbl}
\usepackage{graphicx}
\usepackage{tcolorbox}
\usepackage{arydshln}
\usepackage{algpseudocode}  
\usepackage{caption}

\AtBeginDocument{%
  \providecommand\BibTeX{{%
    \normalfont B\kern-0.5em{\scshape i\kern-0.25em b}\kern-0.8em\TeX}}}

\usepackage{amsmath,amssymb}
\usepackage{comment}

\definecolor{dark-red}{RGB}{255,0,0}
\definecolor{dark-green}{RGB}{0,200,0}

\newboolean{showcomments}
\setboolean{showcomments}{true}
\ifthenelse{\boolean{showcomments}}

\begin{document}

\thispagestyle{plain}
\pagestyle{plain}

\title{The Devil is in the Tails: How Long-Tailed Code Distributions Impact Large Language Models}

\makeatletter
\newcommand{\linebreakand}{%
  \end{@IEEEauthorhalign}
  \hfill\mbox{}\par
  \mbox{}\hfill\begin{@IEEEauthorhalign}
}
\makeatother

\author{
    \IEEEauthorblockN{Xin Zhou\IEEEauthorrefmark{2}, Kisub Kim\textsuperscript{*}\thanks{*Corresponding author. Email: kisubkim@smu.edu.sg}\IEEEauthorrefmark{2}, Bowen Xu\IEEEauthorrefmark{2}\IEEEauthorrefmark{3}, Jiakun Liu\IEEEauthorrefmark{2}, DongGyun Han\IEEEauthorrefmark{4},  David Lo\IEEEauthorrefmark{2}}
    \IEEEauthorblockA{\IEEEauthorrefmark{2}\textit{Singapore Management University, Singapore}
    \\\{xinzhou.2020, bowenxu.2017\}@phdcs.smu.edu.sg, \{kisubkim, jkliu, davidlo\}@smu.edu.sg}
    \IEEEauthorblockA{\IEEEauthorrefmark{3}\textit{North Carolina State University, USA}
    \\bxu22@ncsu.edu}
    \IEEEauthorblockA{\IEEEauthorrefmark{4}\textit{Royal Holloway, University of London, UK}
    \\donggyun.han@rhul.ac.uk}
}

\maketitle

\begin{abstract}\label{abstract}
Learning-based techniques, especially advanced Large Language Models (LLMs) for code, have gained considerable popularity in various software engineering (SE) tasks. However, most existing works focus on designing better learning-based models and pay less attention to the properties of datasets. Learning-based models, including popular LLMs for code, heavily rely on data, and the data's properties (e.g., data distribution) could significantly affect their behavior. We conducted an exploratory study on the distribution of SE data and found that such data usually follows a skewed distribution (i.e., long-tailed distribution) where a small number of classes have an extensive collection of samples, while a large number of classes have very few samples. We investigate three distinct SE tasks and analyze the impacts of long-tailed distribution on the performance of LLMs for code. Our experimental results reveal that the long-tailed distribution has a substantial impact on the effectiveness of LLMs for code. Specifically, LLMs for code perform between 30.0\% and 254.0\% worse on data samples associated with infrequent labels compared to data samples of frequent labels. 
Our study provides a better understanding of the effects of long-tailed distributions on popular LLMs for code and insights for the future development of SE automation.

\end{abstract}
\section{Introduction} 

Data distribution refers to the way in which a set of data is spread out or distributed across different values.
It is a critical factor for machine learning-based approaches, as it is the presence of repetitive patterns within the data distribution that enables the automated tasks to be performed using machine learning tools~\cite{lecun2015deep,tan2018survey}.

Understanding the data distribution in software engineering (SE) could guide researchers to better design automatic tools to help developers.
Hindle et al.~\cite{HindleBSGD12} reported a well-known finding on data distribution namely the naturalness of code: code exhibits a high degree of repetition, which makes code predictable by language models. This fundamental fact supports researchers in employing language models in various SE tasks~\cite{jiang2021cure,svyatkovskiy2020intellicode,t5_review,mularec} to automatically generate/predict code.
Beyond this important finding, we further observe that the data distribution of SE datasets could be long-tailed: some specific code tokens, APIs, libraries, or tools could massively occur in many software systems while a vast number of others only have few occurrences.

Examining this long-tailed distribution is crucial to the software engineering domain not only because of its prevalence in real-world datasets~\cite{shirky2003power,brynjolfsson2003consumer,imagenet-lt} but also its substantial impact on the performance of machine learning models~\cite{yang2022survey_lt}.
An example of the long-tailed distribution in the software engineering domain can be observed in the distribution of Common Weakness Enumeration (CWE)~\cite{treevul}.
Figure~\ref{fig:cwe_type} demonstrates the distribution of security patches across various CWE types gathered from 1,560 open-source software repositories~\cite{treevul}.\footnote{Figure~\ref{fig:cwe_type} only presents the 20 most frequent CWE types and the remaining 93 types with lower frequency are omitted for better visualization.} 
We observe that the top 6 out of 113 CWE types contain almost 50\% of the samples while the remaining 107 types of CWE have fewer occurrences each.
Certainly, a CWE type with low occurrence does not mean it is less important. For instance, 14 out of the top 25 most dangerous CWE types, in the year of 2022~\cite{cwe2022}, were categorized as a part of the 107 infrequent CWE types. An unidentified vulnerability potentially brings critical issues to individuals or organizations when it is exploited by attackers~\cite{dowd2006art,turner2008symantec}.

\begin{figure}
	\centering
	\includegraphics[width=\columnwidth]{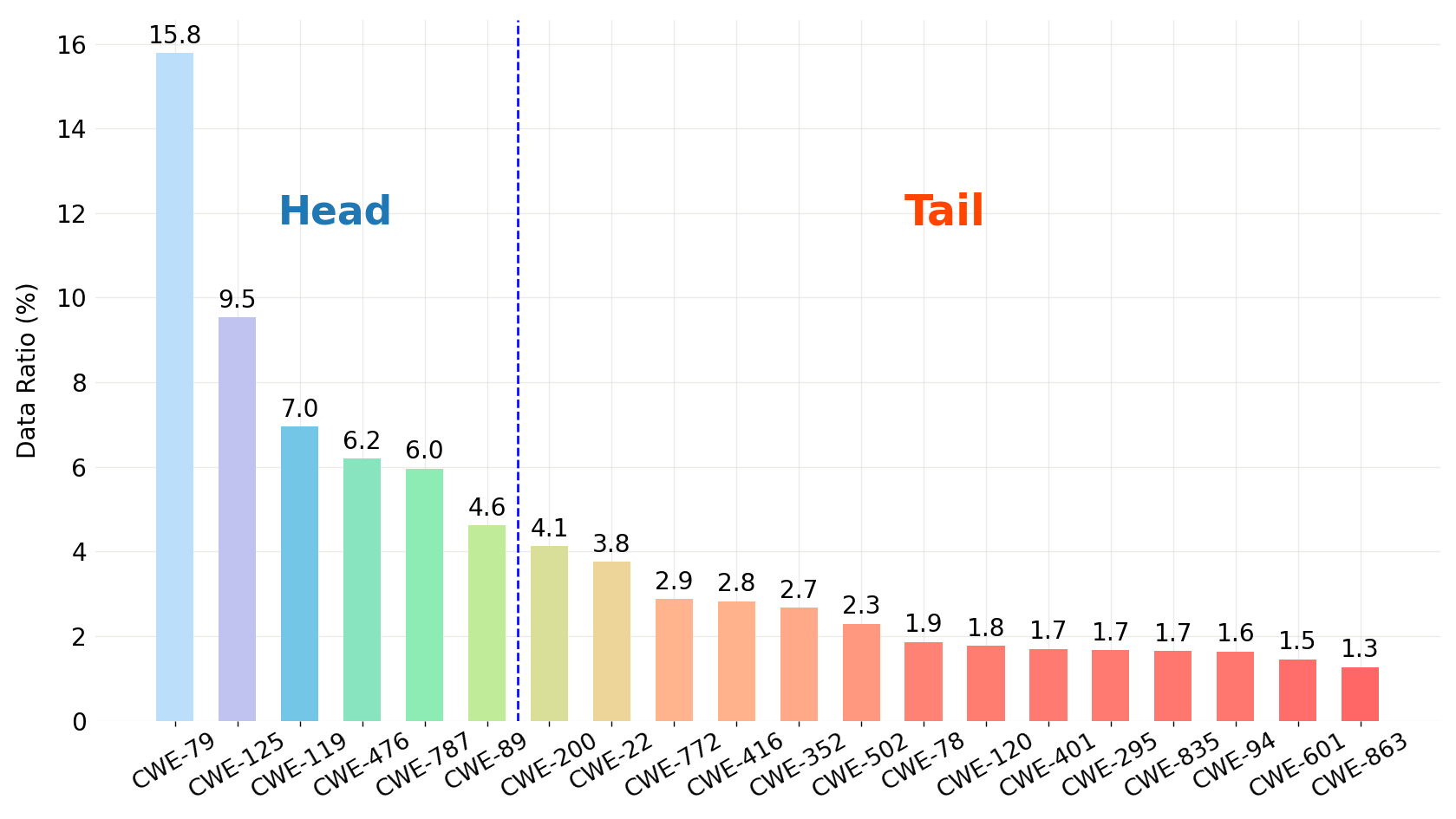}
	\caption{CWE type distribution of security patches~\cite{treevul}.} 
	\label{fig:cwe_type}
 \vspace{-0.5cm}
\end{figure}

In this study, we focus on investigating the existence of long-tailed distributions in labels of SE datasets, as learning-based approaches with labeled data have become the mainstream for most SE tasks~\cite{sharma2021survey,watson2022systematic}. 
Labels in SE datasets are vital as they represent the intents and goals of the tasks~\cite{elahi2010vulnerability}. 
Formally, the \textbf{long-tailed distribution exists in the labels of SE datasets} if \textit{a small proportion of labels has a vast number of data samples, while a large number of other labels have very few data samples.}

To facilitate our analysis, we further categorize all labels into two categories: \textbf{head labels}, which refer to the most frequently occurring labels accounting for 50\% of the data samples, and \textbf{tail labels}, which consist of infrequent labels and account for the remaining 50\% of the data samples. 
For instance, when considering CWE types as labels, the top 6 CWE types depicted in Figure~\ref{fig:cwe_type} correspond to the head labels, while the other 107 CWE types (including 93 omitted types) correspond to the tail labels according to the portion of their data samples.

Without a thorough exploration of the long-tailed distribution, we may not fully realize the poor performance of learning-based approaches on the tail.
A motivating example in Table~\ref{tab:cwe_error} demonstrates the vulnerability type prediction results of the state-of-the-art approach, TreeVul~\cite{treevul}. 
This demonstrates the accuracy scores on patches with the infrequent type ``CWE-863'' (i.e., a tail label), which accounts for only 1.3\% of all patches in the entire dataset. 
Notably, TreeVul achieves a significantly low accuracy (11.1\%) on this specific type compared to the overall accuracy on the entire dataset (73.1\%). 
In contrast, for the frequent type ``CWE-79'' (i.e., a head label), TreeVul achieves a much higher accuracy (88.5\%).

To comprehensively investigate the impact of long-tailed code distribution on learning-based SE models, we focus on two specific categories of SE models: 1) popular Large Language Models (LLMs) for code including CodeBERT~\cite{CodeBERT} and CodeT5~\cite{CodeT5}, and 2) the state-of-the-art approaches of SE tasks. 
Our focus on LLMs for code is based on their widespread popularity and extensive adoption within the SE community~\cite{DBLP:conf/icse/JiangSHSSLTD23,DBLP:conf/icse/NiuLNCGL23,zhou2021assessing}.
A wide range of research endeavors aimed at enhancing the performance of LLMs for code~\cite{DBLP:conf/icse/LiuWXML23},
delving into their robustness~\cite{yang2022natural}, and unveiling the inherent operational mechanisms of LLMs for code~\cite{DBLP:conf/kbse/LopezWCS22}. Moreover, other studies explored techniques for compact model compression~\cite{DBLP:conf/kbse/Shi0XK022} and scrutinizing the effects of data quality issues (e.g., label correctness) on the effectiveness of LLMs for code~\cite{DBLP:conf/icse/CroftBK23}.
However, it is noteworthy that the SE community has not yet substantially investigated to understand the impact of the long-tailed code distribution. 
This underpins our choice to consider LLMs for code as the studied models.
Additionally, we studied the state-of-the-art (SOTA) approaches to distinguish the impacts of long-tailed distribution on the most advanced approaches for SE tasks. Interestingly, the SOTA approaches of the studied tasks are all built upon LLMs.

\begin{table}[t]
\caption{Results of TreeVul on patches of CWE-863 (one infrequent type) and CWE-79 (one frequent type)}
\resizebox{0.48\textwidth}{!}{%
\begin{tabular}{l|c|c}
\hline
\multicolumn{1}{c|}{\textbf{CWE Type}}                                                                         & \textbf{Data Ratio}   & \textbf{Accuracy}          \\ \hline
CWE-863: Incorrect Authorization                                                                               & 1.3\%                 & \multicolumn{1}{c}{11.1\%} \\ \hline
\begin{tabular}[c]{@{}l@{}}CWE-79: Improper Neutralization \\ of Input During Web Page Generation\end{tabular} & 15.8\%                & \multicolumn{1}{c}{88.5\%} \\ \hline 
\end{tabular}
}
\label{tab:cwe_error}
\vspace{-0.5cm}
\end{table}

In this study, we investigate the long-tailed distribution in labeled SE datasets, focusing on three valuable downstream tasks.
We design an automatic long-tailed distribution analysis tool namely LTAnalyzer. LTAnalyzer first identifies the suitable unique labels for both classification and generation tasks and quantifies the long-tailedness based on the Gini coefficient~\cite{gini1912variabilit}. We use LTAnalyzer to identify the existence of long-tailed distributions in the datasets of three valuable downstream tasks.
To understand the impact of long-tailed distribution, we compare the model performance on the head and tail data for the state-of-the-art approaches and popular LLMs for code such as CodeBERT and CodeT5. We observe that those models perform 30.0--254.0\% worse on the tail data compared to the head. We explore the potential solutions (i.e., Focal Loss~\cite{focal} and LRT~\cite{lrt_a}) to improve the performance on the tail, which are originally proposed in computer vision. Specifically, these solutions assign higher weights to the tail during training so that learning-based models may learn more effective features of the tail. The solutions could improve the model performance on the tail images by 9.0\% and 29.0\%, respectively. However, they are relatively ineffective in SE datasets, resulting in only marginal improvements in accuracy scores ranging from 0.3\% to 1.4\%.
Alternatively, we further explore with a simple approach
to identify the tail data during inference (the labels are unknown).
Specifically, we fine-tune the CodeBERT~\cite{CodeBERT} model to perform binary classification (i.e., predicting whether test data belongs to the tail or the head).
The generated predictions of automatic SE tools on the tail data are more likely to be wrong and not reliable. We aim to identify the tail data and warn users about the potential unreliability of the predictions generated by SE tools because they are made on the tail data.
Results show that the simple approach could achieve an accuracy of 66.7\% to 84.4\% in identifying the tail in studied tasks.

We structure our study by answering the following research questions:
\begin{enumerate}[leftmargin=*]
\item  \textbf{RQ1. To what extent do long-tailed distributions exist in the studied SE datasets?}
\item  \textbf{RQ2. How do the long-tailed distributions affect the effectiveness of learning-based SE techniques?} 
\item  \textbf{RQ3. How effective are the potential solutions to address long-tailed distributions for SE tasks?}
\item  \textbf{RQ4. How accurately can we identify the tail?}
\end{enumerate}

Based on our findings, we provide insightful implications for future research:
(1) Researchers should explicitly report the performance of their approaches in tails. (2) New approaches are needed to address the long-tailed distribution. Researchers should consider taking the rich label relationships in SE datasets into account.
(3) Identifying the tail data is a promising direction to mitigate the negative impacts of long-tailed distribution.

In summary, our contributions are as follows:
\begin{itemize}[leftmargin=*]
\item [$\bullet$] To the best of our knowledge, we are the first to study the impact of long-tailed code distribution on LLMs for code.
\item [$\bullet$] We found that the long-tailed distribution has a substantial impact on the performance of LLMs for code.
\item [$\bullet$] We developed an automated tool that can detect and quantify the long-tailed distribution's existence and degree.
\end{itemize}

\section{Preliminaries and Related Work}
\label{sec:background}

In this section, we clarify the definitions of long-tailed distributions in various communities. 
Then, we describe the related work and clarify the differences.

\subsection{Long-tailed Distribution}

It is worth noting that the ``long-tailed distribution'' has different focuses in the business/statistics community and the computer science community. In the business/statistics communities~\cite{anderson2006long,brynjolfsson2003consumer}, long-tailed distribution is defined based on mathematical forms, such as exponential distributions, and is mainly used for theoretical analysis, such as examining whether a business sale distribution follows a specific exponential distribution or not. In contrast, the computer science community uses long-tailed distribution as a general concept to indicate skewed distributions~\cite{yang2022survey_lt,imagenet-lt}.  
In other words, a ``long-tailed distribution'' is synonymous with a ``skewed distribution'' when many classes of labels are available.
The computer science community focuses mainly on analyzing the impact of long-tailed distributions on machine learning models. The majority of studies exploring the impact of long-tailed distributions on machine learning models are conducted in the field of computer vision. Within this domain, researchers mainly focus on image classification tasks~\cite{yang2022survey_lt,lrt_a} and have discovered that machine learning models face difficulties in accurately predicting labels for images in the tail~\cite{yang2022survey_lt}.
In this study, we adopt the computer science community's definition and usage of the term ``long-tailed'' to define and analyze the long-tailed distribution in software engineering data. 
Different from the studies in computer vision, our study investigates classification and generation tasks in SE.

Furthermore, in the realm of SE, some existing studies~\cite{DBLP:conf/oopsla/LopesO15,borges2016understanding} have also noticed the long-tailed distribution phenomenon in code. For instance, Lopes and Ossher~\cite{DBLP:conf/oopsla/LopesO15} unveiled that the sizes of Java projects conform to a long-tailed distribution: a considerable number of Java projects belong to the small or medium scale category, while only a limited portion of these projects are of large magnitude. Similarly, Borges et al.~\cite{borges2016understanding} identified that the popularity pattern of software artifacts exhibits a long-tailed distribution, signifying that a small fraction of software artifacts can attain widespread popularity.
However, those existing studies mainly described their observation of long-tailed code distribution in a specific task/dataset. In contrast, our study first studies the existence of long-tailed code distributions in three distinct SE tasks that have not been explored in prior studies.
Subsequently, we investigate how such long-tailed code distributions influence the performance of popular LLMs for code.

\subsection{Class Imbalance in SE} 
The issue of class imbalance in software engineering (SE) is closely relevant to the long-tailed distribution.
The class imbalance problem is a well-known issue in SE, particularly in defect prediction. In defect prediction, there are naturally many more non-defective instances than defective ones~\cite{DBLP:journals/tse/FentonO00,DBLP:journals/tr/WangY13,DBLP:journals/ese/KameiFMYUH16}. 
Class imbalance negatively impacts the accuracy of defect prediction models~\cite{DBLP:conf/icse/CabralMSM19,DBLP:journals/infsof/Ozturk17,DBLP:journals/infsof/FengKYXBKZ21}. To address this, data oversampling techniques such as random oversampling~\cite{DBLP:conf/msr/HoangDK0U19} and SMOTE~\cite{DBLP:conf/icse/TanTDM15,DBLP:journals/ese/JiarpakdeeTT20,DBLP:journals/tse/BenninKPMM18} are widely employed to generate more instances for the minority classes. Random oversampling increases the number of samples in the minority class by randomly repeating samples from that class, while SMOTE generates new samples for the minority class through linear interpolation between minority neighbors. Besides, SMOTE is mainly used to generate new samples for hand-crafted feature datasets~\cite{DBLP:conf/icse/TanTDM15,DBLP:journals/ese/JiarpakdeeTT20,DBLP:journals/tse/BenninKPMM18}.

Class imbalance typically refers to an uneven distribution of data among a few classes, often only two. In contrast, a long-tailed distribution represents an extreme case of class imbalance where there are numerous classes or labels, creating a more challenging problem.
For instance, computer vision datasets~\cite{imagenet-lt,van2018inaturalist} have 365--8,148 classes, while SE datasets~\cite{mularec,t5_review,treevul} can have 117--99,317 classes, as confirmed later in this paper.
Therefore, the long-tailed distribution requires more sophisticated designs to address a vast number of tail classes/labels with rare data samples.

In this study, we intentionally avoid investigating the techniques of random oversampling and SMOTE for handling class imbalance in the context of long-tailed distribution. This decision is grounded in practical considerations. While random oversampling entails duplicating minority class samples to achieve a balanced dataset among all classes, its applicability diminishes significantly under long-tailed distributions. To illustrate, envision a scenario with 500 classes, where one head class possesses 1,000 samples, and the remaining 499 classes each consist of only 5 samples. Implementing random oversampling would yield a dataset containing 500,000 samples, magnifying its size by approximately \textit{142} times in comparison to the original dataset. This substantial inflation of the dataset would notably hamper the training process of SE models and necessitate substantially greater computational resources. Given that one of our studied datasets encompasses 99,317 classes with at least 5,000 occurrences for head classes, the random sampling technique becomes unfeasible.
Additionally, we notice that the SMOTE technique is tailored for hand-crafted features. However, the input code data in our studied tasks consists of non-structural source code snippets, thereby precluding the direct application of SMOTE to generate new code samples for tail classes.

\section{Methodology}
\label{sec:setup_and_results}

\begin{figure}[t]
	\centering
	\includegraphics[width=1.0\columnwidth]{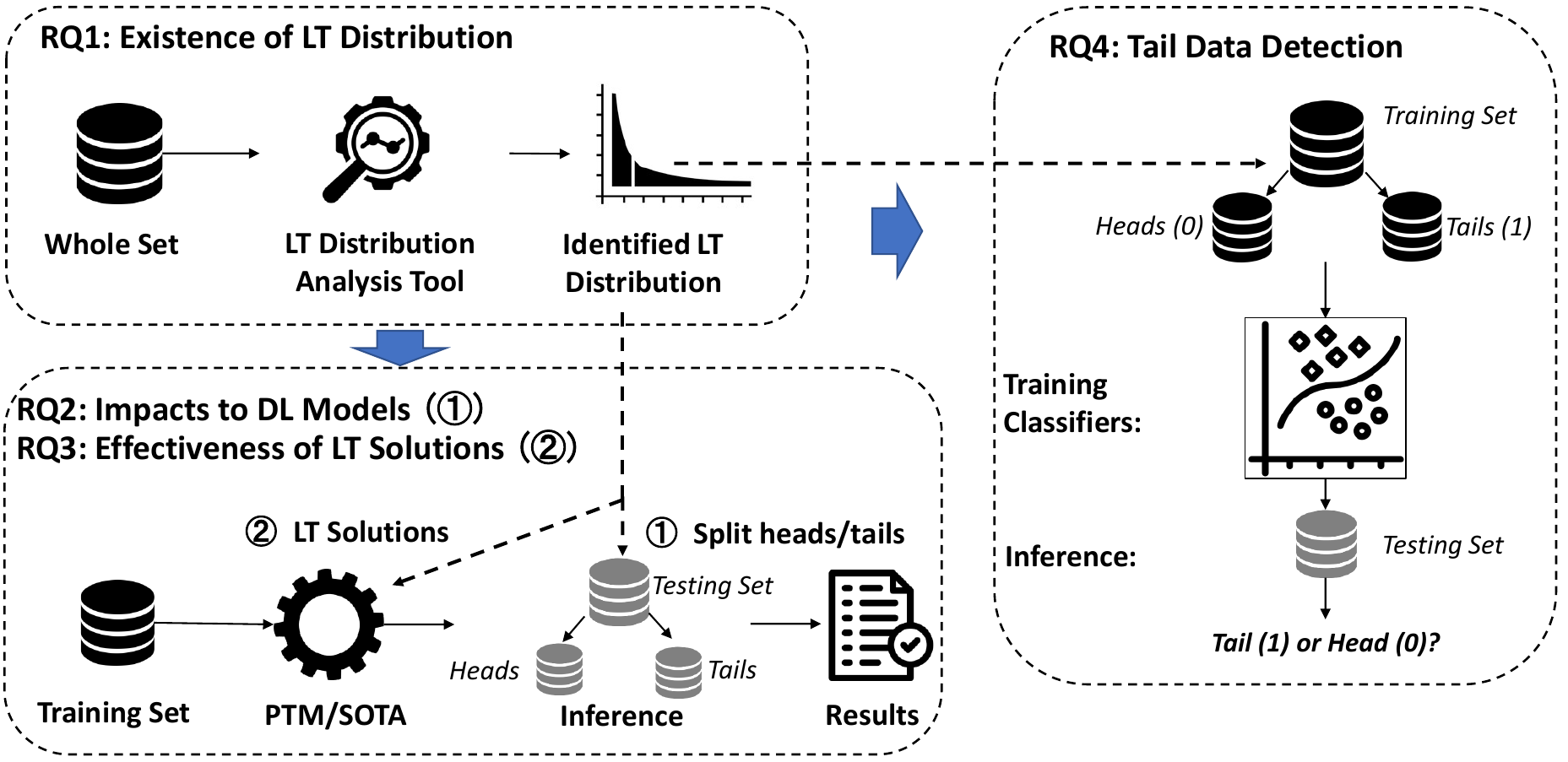}
	\caption{Overview of our research methodology.}
	\label{fig:framework}
\end{figure}

\subsection{Research Questions}
Figure~\ref{fig:framework} provides an overview of our research questions and their interrelationships.
Formally, our study focuses on addressing the following research questions:
\begin{enumerate}[leftmargin=*]
\item  \textbf{RQ1. To what extent do long-tailed distributions exist in the studied SE datasets?}
Long-tailed distributions have been observed in various types of data, such as images and business sales. However, the presence and extent of long-tailed distributions in SE datasets have not been explored yet. Therefore, to investigate this research question, we utilize our proposed long-tailed distribution analysis tool LTAnalyzer to confirm the existence of long-tailed distributions in the datasets of three downstream SE tasks: API sequence recommendation, code revision recommendation, and vulnerability type prediction.

\item  \textbf{RQ2. How do the long-tailed distributions affect the effectiveness of learning-based SE techniques?} 
The impact of long-tailed distributions on learning-based SE techniques has not been thoroughly explored yet. In this research question, we investigate the impact of long-tailed distributions on the effectiveness of learning-based SE techniques, with a focus on two popular LLMs for code (CodeBERT and CodeT5) and state-of-the-art approaches for three studied tasks.

\item  \textbf{RQ3. How effective are the potential solutions to address long-tailed distributions for SE tasks?}
Several techniques have been proposed in computer vision to specifically address long-tailed distributions in images~\cite{yang2022survey_lt,focal,lrt_a}. Although these techniques were not initially developed for SE datasets, they have the potential to be effective solutions to address the long-tailed distributions in SE datasets. Thus, in this research question, we investigate the effectiveness of two techniques (Focal Loss~\cite{focal} and LRT~\cite{lrt_a}) from computer vision on long-tailed SE datasets.
Focal Loss is a widely used solution to mitigate the effect of long-tailed distribution in general. LRT is the state-of-the-art mitigation solution in computer vision.

\item  \textbf{RQ4. How accurately can we identify the tail?}
Learning-based SE approaches may struggle to learn effective features for tail data, which can lead to erroneous predictions. 
During inference, we do not know whether the test data belongs to the head or tail while SE tools are more likely to make wrong predictions on the tail than the head.
To address this, identifying tail data during inference can be useful for warning users about the reliability of predicted labels. Therefore, this research question aims to investigate how accurately we can classify whether a test data instance belongs to the head or tail with a fine-tuned CodeBERT model.
\end{enumerate}

\subsection{Study Tasks}
According to a recent literature review~\cite{watson2022systematic}, the three most commonly studied task formulations in software engineering between 2014 and 2019 were generation-based, classification-based, and retrieval-based tasks.
In this study, we prioritize the top two actively studied task formulations: tasks based on generation and classification.

To select suitable generation and classification tasks to study the extent and impact of long-tailed distribution, we follow two considerations: (1) A candidate task should have a dataset mined from a considerable number of software projects. This enhances the representativeness of the dataset. (2) A candidate task contains diverse labels with a relatively large number of unique labels. 
Following the considerations, we choose three downstream tasks as the study subjects: two generation-based tasks, which are API sequence recommendation~\cite{mularec} and code revision recommendation~\cite{t5_review}, along with a multi-class classification-based task, vulnerability type prediction~\cite{treevul}.

\vspace{0.2cm}
\noindent\textbf{API Sequence Recommendation~\cite{mularec}} is a generation (\textit{sequence-to-sequence}) task to generate API sequence recommendations for the latter part of the code based on the input of natural language (NL) queries and the prior source code contexts.
We utilized a large-scale dataset recently released by Irsan et al.~\cite{mularec}.
This dataset is built on a Java dataset containing 50,000 compilable Java projects provided by Martins et al.~\cite{martins201850k}.
Irsan et al.~\cite{mularec} extracted all the methods from these projects using the Eclipse JDT parser and then extracted the method declaration, method body, and Javadoc annotation from each method. They further extracted the API sequence from the method body using the same parser.
For the inputs and outputs, Irsan et al. defined the NL query as the first sentence of the Javadoc annotation, 
the code context as the method declaration and the first three lines of the method's source code, and the ground truth API sequence as the API sequence extracted from the source code after the first three lines. 
In addition, there are a large number of (i.e., 99,317) unique APIs available in the dataset shared by Irsan et al.~\cite{mularec}.

\vspace{0.2cm}
\noindent\textbf{Code Revision Recommendation} is another \textit{sequence-to-sequence} task where the DL model takes bi-modal inputs ($code$ and $comments$), and generates the $revised \, code$. The comments are provided by the code reviewers, which explain the necessary improvements toward the originally submitted $code$. The $revised \, code$ is the revised version of the submitted code to address the comments.  
We utilized a large-scale dataset released by Tufano et al.~\cite{t5_review}. The dataset was constructed from two sources: 1) Github, comprising open-source projects (4,901 projects) using the web application shared by Dabic et al.~\cite{dabic2021sampling}, and 2) Gerrit, consisting of code review data about 6,388 projects from six Gerrit installations. Tufano et al. extracted triplets $\langle code, comments, revised \, code \rangle$ from both the GitHub and the Gerrit projects. 
Furthermore, the ground truth code revisions in this dataset are diverse as they address comments from different reviewers across a large number of projects and the reviewers could have various intentions when requesting a code revision~\cite{ebert2018communicative}.

\vspace{0.2cm}
\noindent\textbf{Vulnerability Type Prediction} is a multi-label classification task that aims to automatically predict the fine-grained CWE category for an input security patch (i.e., a commit) which can reduce the burden of manual analysis and categorization by human experts. 
Notably, the CWE categories are organized in a tree-like hierarchy, and the level of abstraction in the predicted CWE category is determined by a given depth parameter $d$, which corresponds to the predicted CWE category at the $d$-th level of the CWE tree. 
Pan et al.~\cite{treevul} defined three different levels of target depths: 1, 2, and 3, where d=3 represents the most fine-grained and accurate level in their study. In our work, we focus on d=3, as it represents the most accurate CWE category for an input security patch.
We used a large security patch dataset shared by Pan et al.~\cite{treevul}, collected from the National Vulnerability Database (NVD), consisting of 6,541 commits from 1,560 GitHub open source software repositories.
In addition, this dataset contains 113 unique CWE types, which covers more CWE types than its prior works~\cite{zhou2021finding,zhou2021spi,chen2020machine}.

API sequence recommendation has continuously been studied in the SE community for the past five years~\cite{gu2016deep,fowkes2016parameter,martin2022deep,mularec}.
Code revision recommendation, as part of code review automation, has garnered increasing attention in two years~\cite{tufano2021towards,thongtanunam2022autotransform,t5_review}. 
Vulnerability type prediction is an emerging task that classifies security patches into fine-grained software vulnerability types, facilitating a better understanding of security patches and supporting early remediation~\cite{treevul}. 
These three tasks cover different stages in the software development and maintenance life cycle (i.e., implementation, code review, and vulnerability analysis), and their labels are also distinct from each other: APIs, code changes, and vulnerability types. 

Please note that we did not choose defect prediction and vulnerability prediction as the studied tasks, although they are popular. We aim to study the long-tailed distribution in the labels of SE datasets, which requires diverse labels in the datasets. Most studies~\cite{DBLP:conf/icse/CabralMSM19,DBLP:journals/infsof/FengKYXBKZ21,zhou2019devign,nguyen2022regvd} on defect/vulnerability prediction aim for binary classification.

\noindent
\subsection{Studied Models}
We investigated two categories of models for each task: 1) two widely used LLMs for code, CodeBERT~\cite{CodeBERT} and CodeT5~\cite{CodeT5} fine-tuned for each task, and 2) the state-of-the-art approach for each task.

CodeBERT is a bimodal pre-trained encoder model that is pre-trained on masked language modeling~\cite{bert} and replaced token detection~\cite{electra} tasks. 
CodeT5 is a pre-trained encoder-decoder model that utilizes multiple pre-training tasks~\cite{CodeT5}, i.e., masked span prediction, masked identifier prediction, and identifier tagging. 
We adopted CodeBERT and CodeT5 in a straightforward way. If there are multiple input sources (e.g., code and texts), we concatenated the input sources and fed the concatenated inputs to the models to predict/generate outputs.

For API sequence recommendation, we incorporate the state-of-the-art MulaRec proposed by Irsan et al.~\cite{mularec}.
MulaRec uses CodeBERT~\cite{CodeBERT} as the backbone model, alongside the introduction of an innovative multi-modal fusion module. This multi-modal fusion module effectively aggregates features from natural language queries and programming language contexts.
For code revision recommendation, we include T5-Review proposed by Tufano et al.\cite{t5_review}. It is pre-trained on 1.4 million code snippets from the CodeSearchNet dataset\cite{codesearchnet} and the text from Stack Overflow~\cite{so_data}.
For vulnerability type prediction, we employ TreeVul invented by Pan et al.~\cite{treevul}. 
TreeVul is built upon CodeBERT~\cite{CodeBERT} and it explicitly leverages the relations of the CWE category with its ancestors in the hierarchy.

\begin{figure*}[t]
    \centering
    \begin{subfigure}[b]{0.33\textwidth}
      \includegraphics[width=1.0\textwidth]{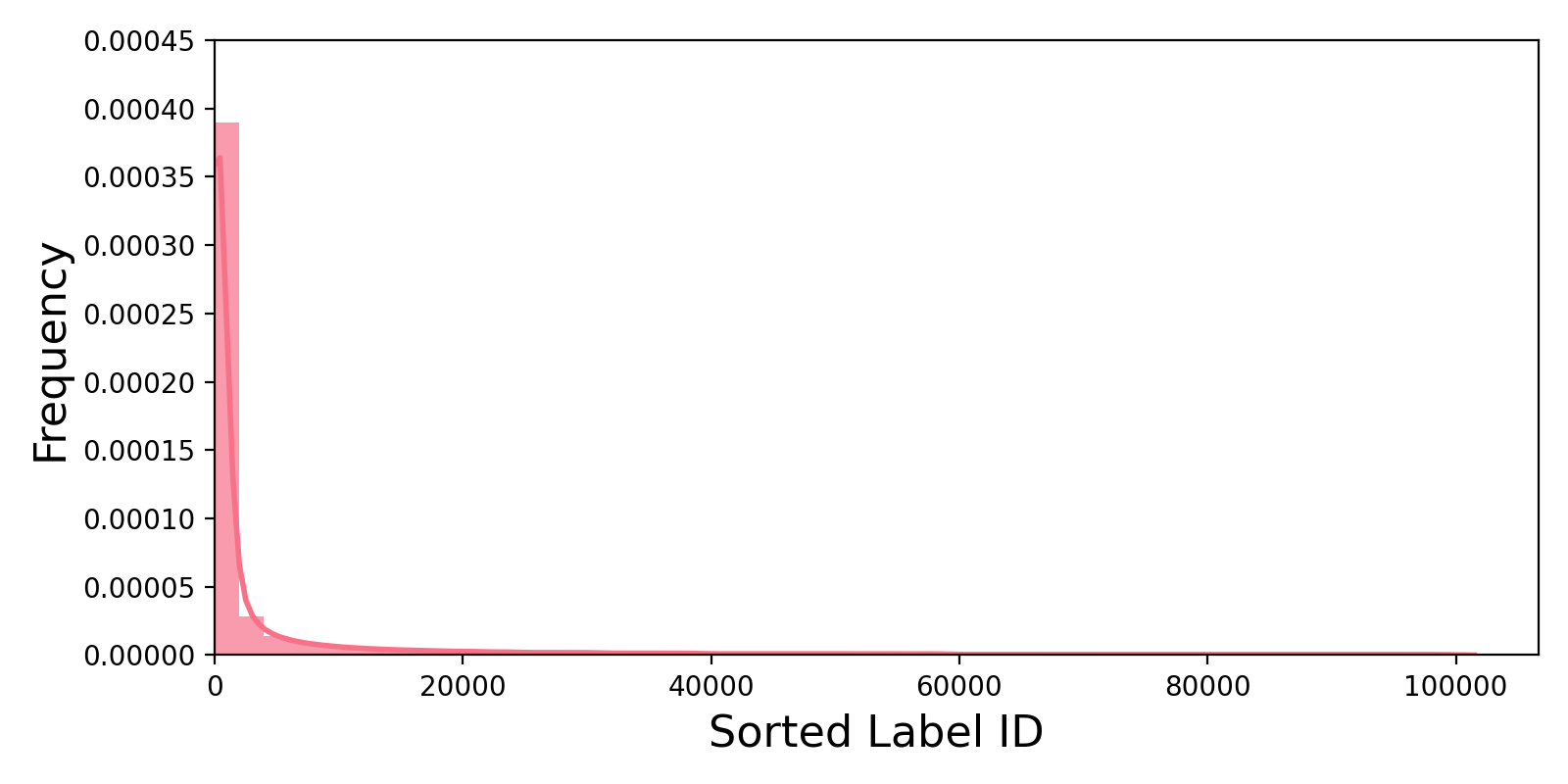}
      \caption{API Distribution}
      \label{fig:api_dis}
    \end{subfigure}%
    \begin{subfigure}[b]{0.33\textwidth}
      \includegraphics[width=1.0\textwidth]{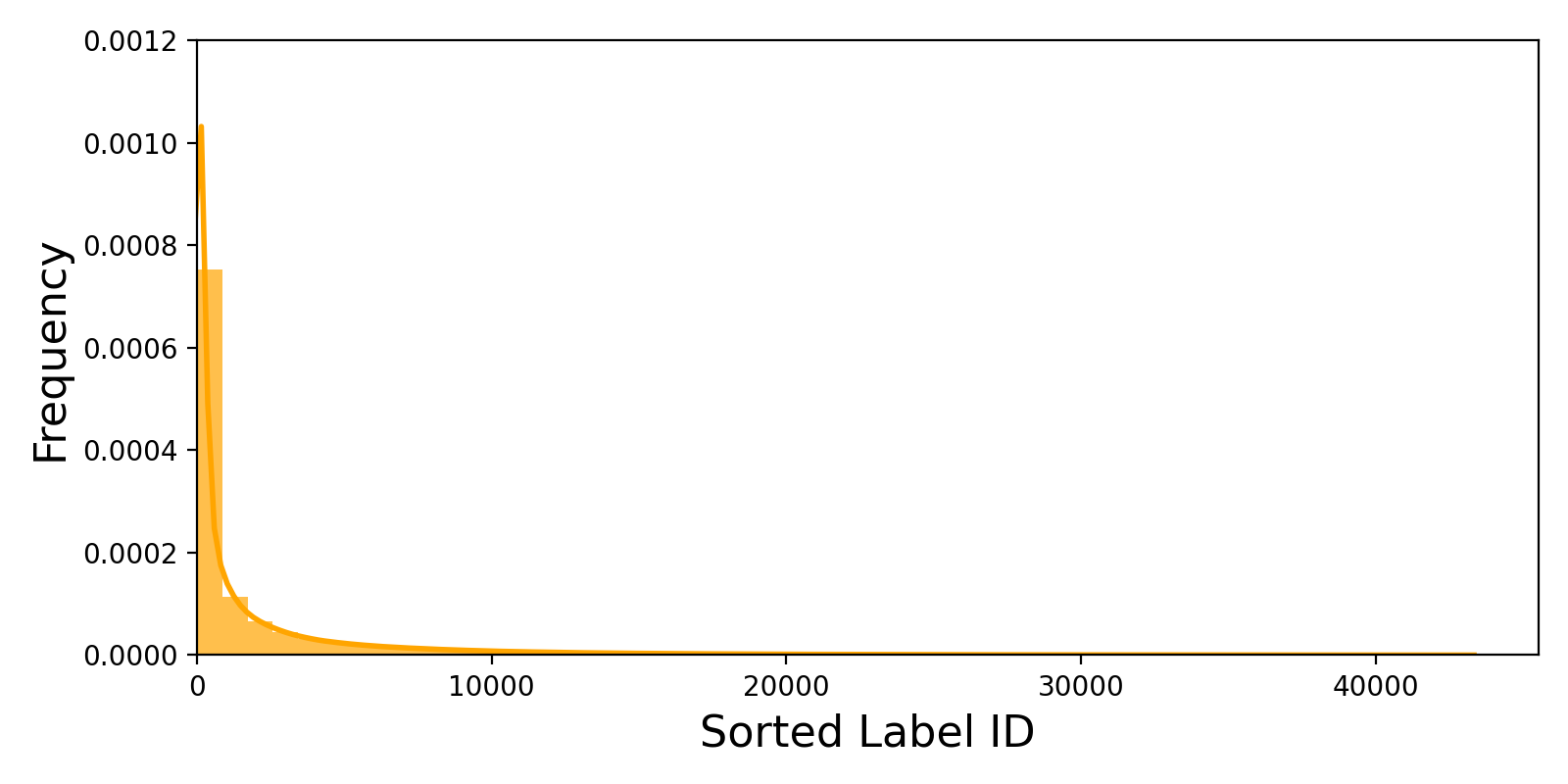}
      \caption{Revision Distribution}
      \label{fig:revision_dis}
    \end{subfigure}
    \begin{subfigure}[b]{0.33\textwidth}
      \includegraphics[width=1.0\textwidth]{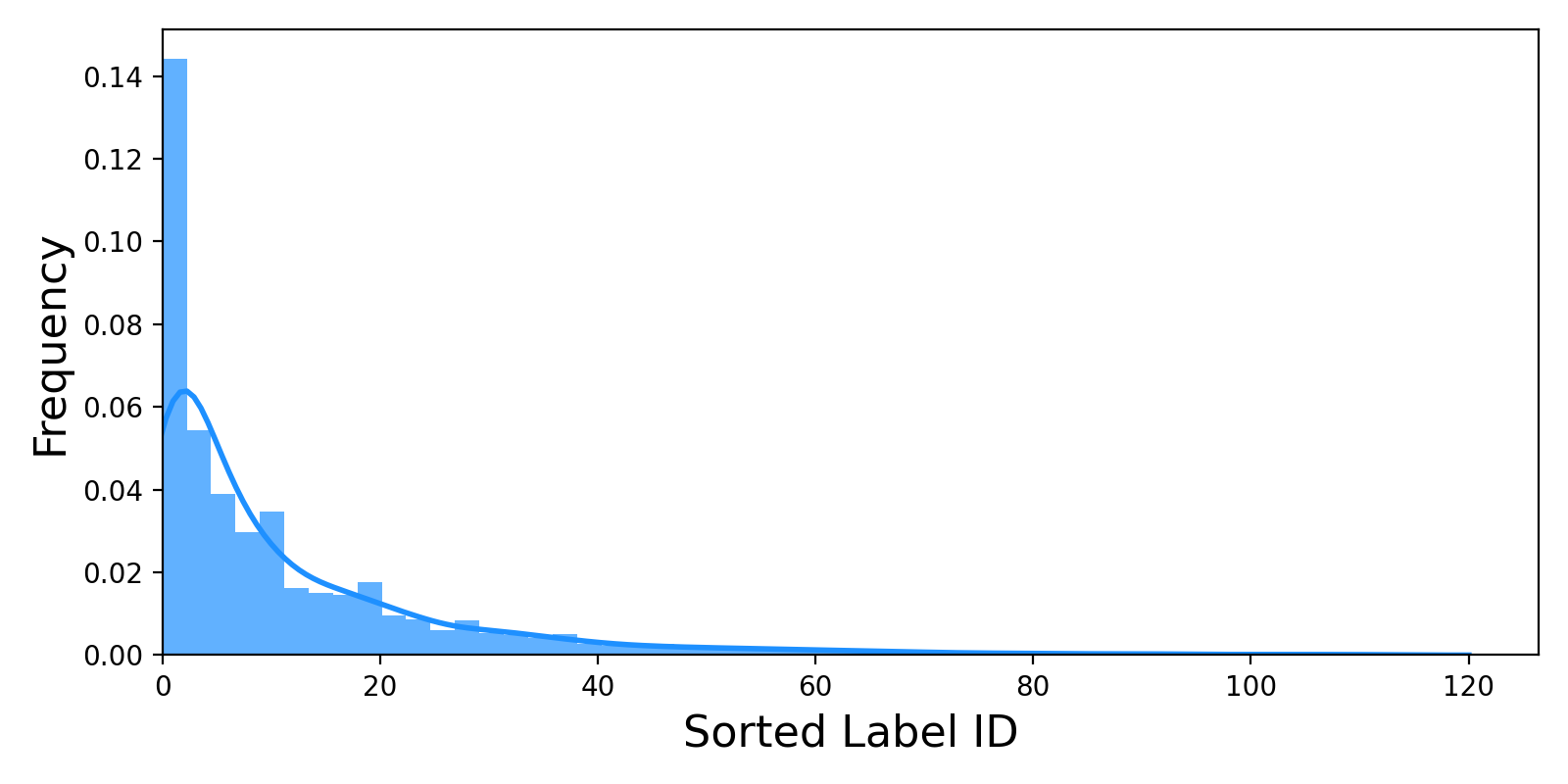}
      \caption{Vulnerability Distribution}
      \label{fig:vul_dis}
    \end{subfigure}%
    \vspace{-0.1cm}
    \caption{Visualization of code distributions in three studied tasks.}
    \label{fig:all_LT_dis}
\end{figure*}

\subsection{Long-tailed Distribution Analysis Tool: LTAnalyzer}
\label{sec:LT}

In order to facilitate the analysis of the long-tailed distribution in a given dataset, we develop a tool named LTAnalyzer which takes a labeled dataset of a task as input and outputs the extent of long-tailedness. This tool enables the quantification of long-tailedness based on the data distribution.

\subsubsection{Analyzing Data Distribution}

The definition of data distribution is given by $P(X)$, where $X$ is the unique label set  and $P(X)$ is the data frequencies of each unique label within the entire dataset. 
It is straightforward to identify the unique labels in classification tasks if the labels are mutually exclusive, meaning that a data sample can only belong to a single label. 

\vspace{0.1cm}
\noindent\textbf{Identifying Unique Labels for Generation Tasks.}
For generation models, there is no clear concept of labels as they employ a ground truth sequence to evaluate the performance while each meaningful token or a set of tokens is used to optimize the model during training. 
Specifically, learning-based models~\cite{sutskever2014sequence} usually generate the output sequence token-by-token instead of a whole sequence at once, indicating that individual tokens may be more appropriate labels for generation tasks.
However, a single token neither establishes meaningful semantics nor represents a data instance. 
To investigate the distribution of the datasets, we define different labels for each generation task by considering the individual characteristics. 
We select the tokens that build complete or independent semantics such as a single API. 
In detail, as an API may be the finest-grained artifact for training the API sequence generation, we define an API as a label for the API recommendation task. 
On the other hand, the revised part of the code in the code revision recommendation plays the key in optimizing the model training. 
We structure the revised code in terms of token-level edits and utilize them as labels (e.g., An edit operation as $\langle$public, ``''$\rangle$ and $\langle$``'', protected$\rangle$) rather than a whole code chunk.
The following cases demonstrate more formalized label identification from generation-based approaches:

\noindent\textbf{Case 1. } In API sequence recommendation, the objective is to suggest \textit{a sequence of APIs}, where a single API is a fundamental unit. Since a single API represents relatively complete functionality, we consider \textit{a single API} as a unique label in this task.

\noindent\textbf{Case 2. } In code revision recommendation, the goal is to suggest the \textit{revised code} that improves the quality of the submitted code following reviewers' comments. Although the ground truth is a sequence of code tokens, the changed code tokens (e.g., newly added or deleted) are specifically more important than those tokens remaining unchanged. To emphasize the changed tokens, we follow prior works~\cite{liu2020automating,shi2022race} to adopt \textit{difflib}~\cite{difflib} to extract the token-level edits from the code changes (input code $\rightarrow$ revised code). A token-level edit is defined as a pair of code tokens $\langle${\tt token before}, {\tt token after}$\rangle$. For instance, if the edit action is {\tt add}, then the {\tt token before} will be an empty string, and the {\tt token after} will be the newly added token in the revised code. In this case, we consider each unique token-level edit as one unique label.

\subsubsection{Quantifying Long-Tailedness} 
Accurate measurement of the long-tailedness of data is an important aspect of understanding the dataset~\cite{yang2022survey_lt}. 
In this work, we adopt the Gini Coefficient~\cite{gini1912variabilit} as the metric to quantify the long-tailedness.
The Gini Coefficient is recommended as an ideal metric by a systematic survey of long-tailed distribution studies in computer vision~\cite{yang2022survey_lt} because it is not affected by the number of data samples in the datasets.

We use $x_i$ to represent the number of data samples that belong to a label $i$.  A dataset contains $n$ labels in total, with an average number of samples of $\bar{x} = \frac{\sum_{i=1}^{n} x_i}{n}$. Then the Gini coefficient is defined as half of the relative mean absolute difference~\cite{gini1912variabilit}, formally represented as  
$$Gini = \frac{ \sum_{i=1}^{n} \sum_{j=1}^{n} |x_i - x_j| }{2n^2 \bar x} $$ 
When the numerator (i.e., $\sum_{i=1}^{n} \sum_{j=1}^{n} |x_i - x_j| $) is large, it means that there are big differences among the number of samples belonging to different labels, indicating a higher long-tailedness.
A Gini coefficient of 0\% denotes perfect equality, where each label has the same number of data samples, while a coefficient of 100\% signifies complete inequality, where one label possesses all the data samples. Therefore, the higher the Gini coefficient, the stronger the degree of long-tailed distribution is.

\section{Experiments} \label{sec:results}
In this section, we present experimental results to answer the research questions one by one.
\subsection{RQ1: Extent of LT Distribution}
\label{sec:rq1}

We study the SE datasets by (1) visualizing the distribution, and (2) quantifying the extent of long-tailedness.

\vspace{0.1cm}
\noindent \textbf{Visualizing the data distribution}.  
Figure~\ref{fig:all_LT_dis} visualizes the data distribution in all of the studied tasks. The labels for each task are sorted in descending order according to their frequency, i.e., $\frac{\#\text{occurrence}\,\text{of}\,\text{a}\,\text{label}}{\#\text{occurrence}\,\text{of}\,\text{all}\,\text{labels}}$. Each label is then assigned a unique label ID ranging from 0 to the total number of unique labels minus one.
As shown in Figure~\ref{fig:all_LT_dis}, the datasets for all three tasks exhibit long-tailed distributions. Only a small fraction of labels have high frequencies, while a broad range of other data points has low frequencies of occurrence.

\vspace{0.1cm}
\noindent \textbf{Numbers of Head and Tail Labels}. 
We count the number of head and tail labels in the studied datasets.
In the API sequence recommendation dataset, the top 213 frequent APIs (out of 99,317 unique APIs) account for half of the data samples.
In the code revision recommendation dataset, the top 272 token-level edits (out of 42,825 edits) account for half of the samples. 
Meanwhile, in the vulnerability type prediction dataset, the top 6 frequent CWE types (out of 113) make up half of the samples. 
In addition, although the tail labels have fewer data samples each, there are many important labels in the tails. For instance, 14 out of the top 25 most dangerous CWE types, in the year of 2022~\cite{cwe2022}, belong to the tails.  
This observation motivates us to analyze the performance of SE models on the tail in the next research question.

\vspace{0.1cm}
\noindent \textbf{Quantifying the extent of long-tailedness}. 
Using the Gini coefficient~\cite{gini1912variabilit}, we have quantified the degree of the long-tailedness present in each dataset, as demonstrated in Table~\ref{table:gini}. A higher Gini coefficient signifies a higher long-tailedness, implying that a greater number of data samples belong to fewer head classes/labels. To gain a deeper understanding of the extent of the long-tailedness in SE datasets, we compared it with popular long-tailed computer vision (CV) data, such as ImageNet-LT~\cite{imagenet-lt}, Places-LT~\cite{imagenet-lt}, iNaturalist 2018~\cite{van2018inaturalist}, and LVIS v0.5~\cite{gupta2019lvis} by using the same metric, Gini coefficients, in Table~\ref{table:gini}. It is noteworthy that SE datasets have Gini coefficients ranging from 0.777 to 0.932, 
while those for CV datasets ranged from 0.524 to 0.825. The results demonstrate that on average our studied SE datasets have higher long-tailedness than the well-known long-tailed datasets. 
This could be due to the fact that the studied SE datasets~\cite{mularec,t5_review,treevul} have a larger number of unique labels (ranging from 113 to 99,317 labels) compared to CV datasets (from 65 to 8,148)~\cite{imagenet-lt,van2018inaturalist,gupta2019lvis}.

\begin{table} [t]
\centering
\caption{Long-tailedness of SE datasets and CV datasets}
\resizebox{0.4\textwidth}{!}{%
\begin{tabular}{c|c|c}
\hline
\textbf{Domains}    & \textbf{Datasets}         & \textbf{Gini Coef.} \\ \hline
\multirow{3}{*}{SE} & API Sequence  Rec.~\cite{mularec}        & 0.907               \\
                    & Code Revision Rec.~\cite{t5_review}        & 0.932               \\
                    & Vulnerability  Type Pred.~\cite{treevul} & 0.777               \\ \hline
\multirow{4}{*}{CV} & ImageNet-LT~\cite{imagenet-lt}               & 0.524               \\
                    & Places-LT~\cite{imagenet-lt}                & 0.671               \\
                    & iNaturalist 2018~\cite{van2018inaturalist}         & 0.620               \\ 
                    & LVIS v0.5~\cite{gupta2019lvis}   & 0.825   \\ \hline
\end{tabular}
}
\label{table:gini}
\vspace{-0.3cm}
\end{table}

\begin{figure*}[t]
    \centering
    \begin{subfigure}[b]{0.31\textwidth}
    \centering
      \includegraphics[width=1.0\textwidth]{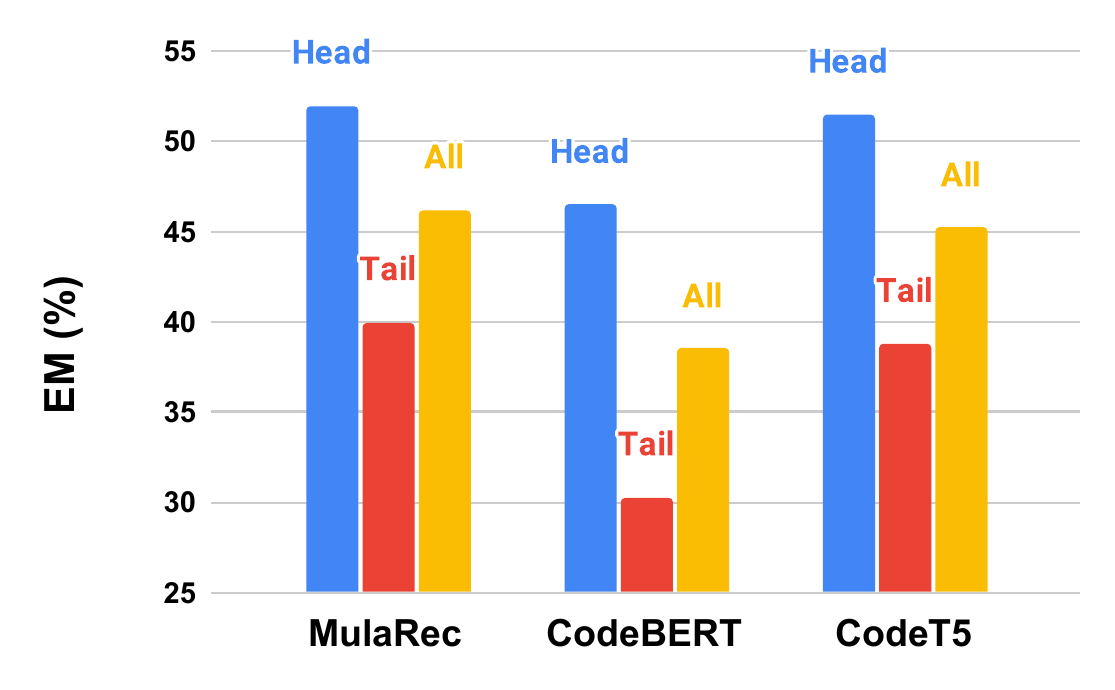}
      \caption{API Sequence Rec.}
      \label{fig:api_impact}
    \end{subfigure}%
    \begin{subfigure}[b]{0.31\textwidth}
    \centering
      \includegraphics[width=1.0\textwidth]{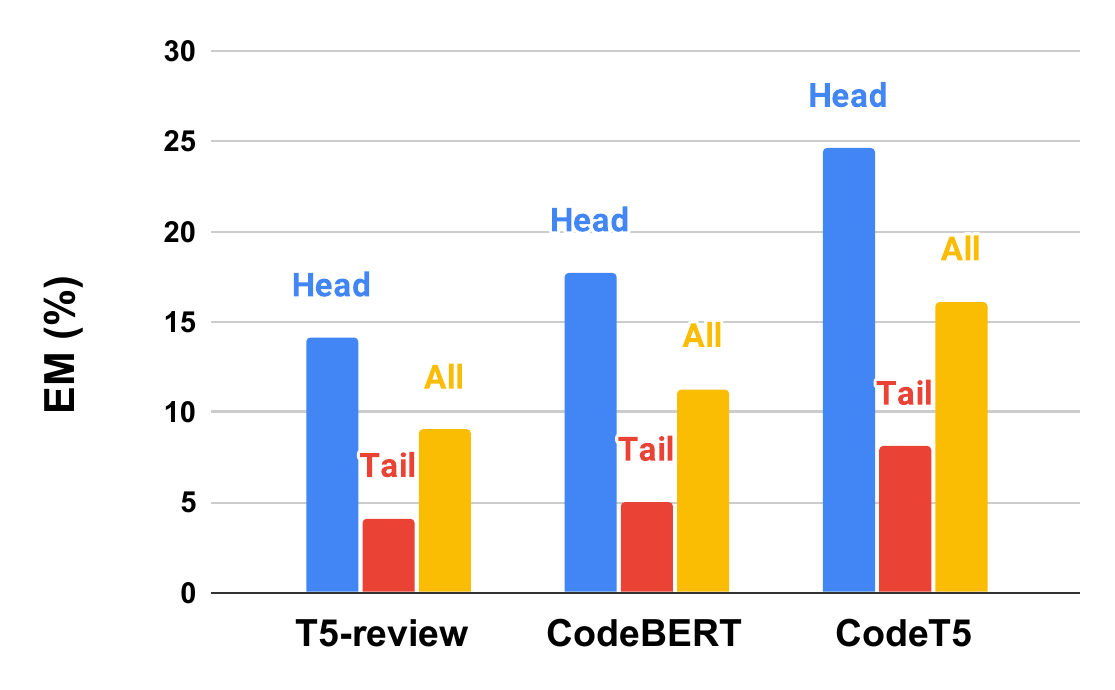}
      \caption{Code Revision Rec.}
      \label{fig:revision_impact}
    \end{subfigure}
    \begin{subfigure}[b]{0.31\textwidth}
    \centering
      \includegraphics[width=1.0\textwidth]{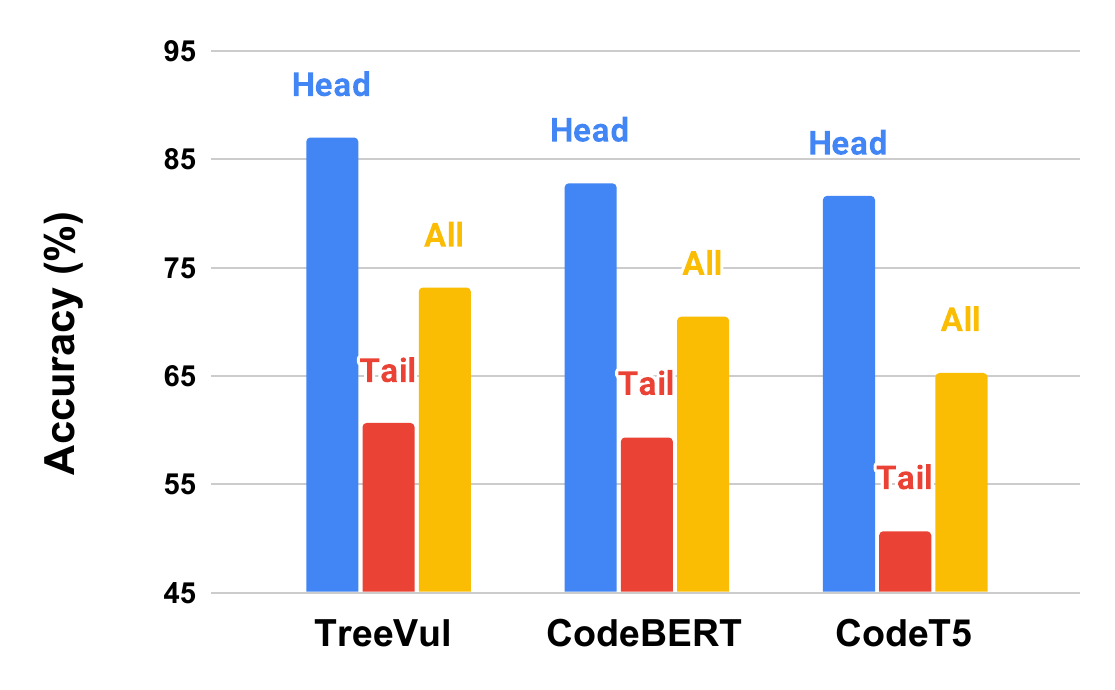}
      \caption{Vulnerability Type Rec.}
      \label{fig:vul_impact}
    \end{subfigure}%
    \caption{Model performances on different subsets of the test datasets in three studied tasks.}
    \label{fig:all_LT_impact}
    \vspace{-0.3cm}
\end{figure*}

\begin{figure*}
    \centering
    \begin{subfigure}[h]{0.3\textwidth}
      \includegraphics[width=1.0\textwidth]{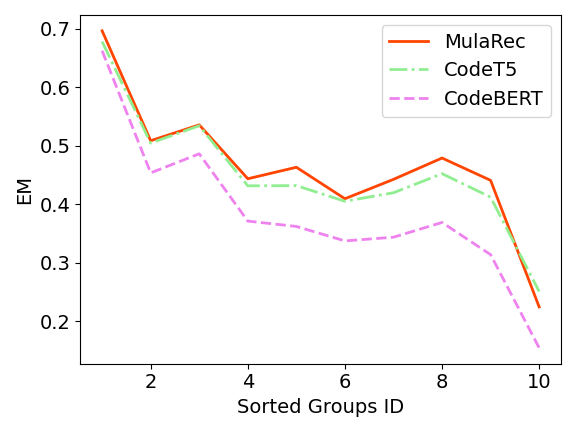}
      \caption{API Sequence Rec.}
      \label{fig:api_impact}
    \end{subfigure}%
    \begin{subfigure}[h]{0.3\textwidth}
      \includegraphics[width=1.0\textwidth]{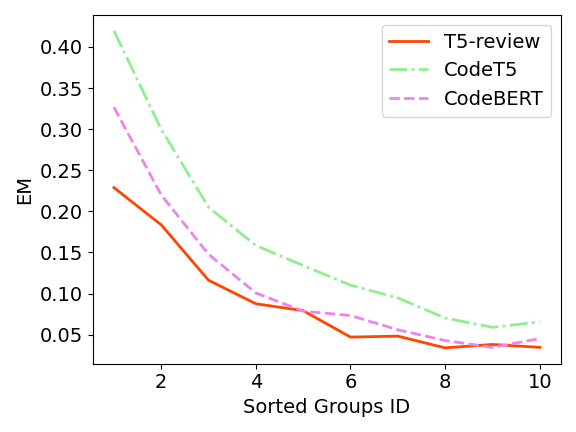}
      \caption{Code Revision Rec.}
      \label{fig:revision_impact}
    \end{subfigure}
    \begin{subfigure}[h]{0.3\textwidth}
      \includegraphics[width=1.0\textwidth]{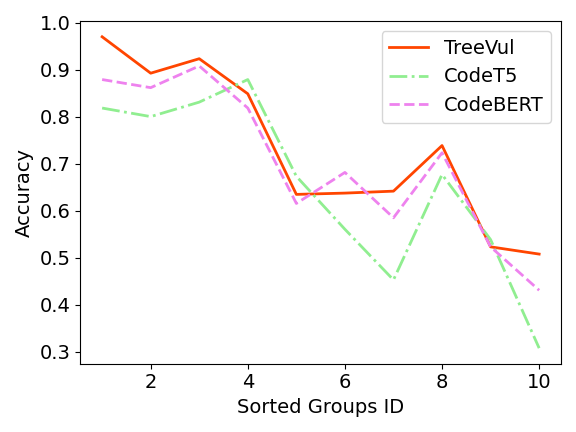}
      \caption{Vulnerability Type Rec.}
      \label{fig:vul_impact}
    \end{subfigure}%
    \caption{Model performances on 10 small groups of the test sets of three studied tasks.}
    \label{fig:all_LT_impact_finegrained}
    \vspace{-0.5cm}
\end{figure*}

\vspace{0.2cm}
\noindent
\begin{tcolorbox} [boxrule=0.8pt,
                top=0.2pt,
                  bottom=0.2pt]
    \textbf{Answer to RQ1}: The datasets of the three studied tasks display different degrees of long-tailed distributions. Within our investigations, some SE datasets may exhibit higher long-tailedness compared to CV datasets. 
\end{tcolorbox}

\subsection{RQ2: Impacts on DL Models}
\label{sec:result2}

In this research question, we investigate the performance of SE models on the tail and head data respectively.

\vspace{0.1cm}
\noindent \textbf{Experimental Setup}.
In order to assess the impact of the long-tailed distribution, we began by splitting the test set equally into two parts: the head and tail \textbf{data} ($\langle$input, ground truth$\rangle$ pairs), based on the frequency of their labels.
The split was performed as follows:

\begin{enumerate}[leftmargin=*]
\item \textit{\textbf{Head Data:}} consisted of 50\% of the data samples whose labels were more frequent.
\item \textit{\textbf{Tail Data:}} comprised of the remaining 50\% of the data samples whose labels were less frequent.
\end{enumerate}
To clarify, if there is a label containing data samples that overlap between the head and tail data, we assign the label to the group that results in the smallest difference between the two groups. For example, assume that there is a label $x$ that contains 10 samples that overlap between the head and tail groups. Prior to adding label $x$ to either group, the head group contains 85 samples and the tail group contains 90 samples. In this case, we assign label $x$ to the head group, since the difference between the two groups after the assignment will be 5, which is smaller than the difference (i.e., 15) 
that would result from assigning label $x$ to the tail group.

For generation-based tasks, ground truth sequences are usually used to evaluate the model performance.
A ground truth sequence may contain multiple labels such as multiple APIs or token-level edits. In such cases, we computed the sum of the reciprocal frequencies of labels in ground truth sequences as the sequence-level score, i.e., $\sum\limits_{x_i\in x_{set}} \frac{1}{freq(x_i)}$, where $freq(.)$ is the frequency of a label (e.g., an API and a token-level edit) in the entire dataset and $x_{set} = [x_0, x_1, ...,x_n]$ refers to the labels contained in the ground truth for a $\langle$input, ground truth$\rangle$ pair. 
We use the reciprocal frequencies of labels to highlight the contributions of infrequent fine-grained labels (tails) in the sequence-level score. 
Then, the head and tail subsets were then identified by analyzing the sequence-level scores: half of the data with higher sequence-level scores (containing more infrequent labels) is the tail data and the rest is the head data.

To analyze the performance of the studied DL models, we use two widely-used metrics accuracy and Exact Match (EM)~\cite{codexglue,t5_review,thongtanunam2022autotransform}. 
Specifically, for the classification task, we adopt accuracy which measures the proportion of correctly classified samples out of the total number of samples. For generation-based tasks, we adopt EM which measures the percentage of cases where the generated answer exactly matches the ground truth. Due to the limited space, we put the results in terms of other widely used metrics (e.g., BLEU score and F1 score) in the appendix of the replication package.

\vspace{0.1cm}
\noindent \textbf{Experimental Results}.
The experimental results for each task are depicted in Figure~\ref{fig:all_LT_impact}, where the blue, red, and yellow bars represent the model performance on the head, tail, and the whole test set, respectively.
In the case of the API sequence recommendation task, we observe that the state-of-the-art model (MulaRec), CodeBERT, and CodeT5, perform 30.0\%, 53.3\%, and 32.7\% better on the head data than on the tail data, respectively.
Similarly, in the code revision recommendation task, T5-Review, CodeBERT, and CodeT5 exhibit 243.9\%, 254.0\%, and 203.7\% better performance on the head data than on the tail data, respectively.
For the vulnerability type prediction task, TreeVul, CodeBERT, and CodeT5 achieve 43.6\%, 39.4\%, and 60.9\% more accurate results on the head data compared to the tail data.

In our previous experiments, we divided the test set into only two large groups based on frequency or sequence-level scores. In addition, we conducted a more thorough analysis by investigating the model performance in more small groups. 
To achieve this, we sorted the test set in descending order with respect to label frequencies for the classification task. For generation tasks, we sorted the test set in ascending order with respect to sequence-level scores, i.e., from sequences containing few infrequent labels to those with many infrequent labels.
Then, we divided the sorted test set into 10 groups, with each group consisting of an equal number (10\% of the data) of data samples.
Afterward, we separately compute the evaluation metrics for the predictions made by a SE model on each of the ten groups of test data. 
The line plot in Figure~\ref{fig:all_LT_impact_finegrained} illustrates the results of the experiment, with the red, green, and pink lines representing the state-of-the-art, CodeT5, and CodeBERT models, respectively.

In the API sequence recommendation task, we found that for groups with the highest frequency (first 10\% of data), most models achieved EM scores higher than 65\%. However, for groups with the lowest frequency (last 10\% of data), the EM scores ranged from 12--25\%. This indicates the model performance drops when the class/label frequencies decrease.
Moving on to the code revision recommendation task, we observed about 3--7\% EM scores in the group with the lowest frequent labels (last 10\% of data). 
This is poor compared to the 23-42\% EM scores obtained for the first 10\% of data.
Regarding the task of vulnerability type prediction, we observe that the impact of long-tailedness was less than that in the preceding two tasks, as evidenced by the absence of a consistent decline in model performance as the class/label frequencies decreased.
Instead, we uncover a distinct staircase pattern in the graph, where the results were better (80--98\% in accuracy) for the more frequent data labels (first 40\% of the data), while the performance significantly dropped (31--73\% in accuracy) for less frequent data labels (last 60\% of the data).

\vspace{0.1cm}
\noindent \textbf{Qualitative Analysis}.
To illustrate the impact of long-tailed distributions, we present one case for each task in Figure~\ref{fig:qualitative}.
Figure~\ref{fig:qualitative}(a) illustrates the input (text annotation and code context), ground truth API sequences, and API sequences generated by MulaRec in the API sequence recommendation. MulaRec successfully generated the first two APIs (\emph{Reader.read} and \emph{System.arraycopy}) that are relatively frequent with 426 and 6,205 data samples in the dataset, respectively. However, it failed to generate the infrequent API \emph{String.copyValueOf} which has only 47 data samples and belongs to the tail. 
Instead, MulaRec generated a wrong API (\emph{String.$<$init$>$}) which contains 5,046 samples.

Figure~\ref{fig:qualitative}(b) presents the input, the ground truth revised code, and the generated code by T5-Review. The ground truth aims to modify the access level of a method from {\tt public} to {\tt protected} with two token-level edits: deleting {\tt public} and inserting {\tt protected} in the corresponding position. 
T5-Review succeeded in deleting {\tt public}, which has 565 data samples. 
However, it failed to insert {\tt protected} due to limited data instances, i.e., only 79 belong to the corresponding label.
Notably, the token-level edit ``insert {\tt protected}'' is one of the tail labels.

In the case of vulnerability type prediction, Figure~\ref{fig:qualitative}(c) exhibits the predictions of TreeVul on patches of type CWE-661, which make up only 1.1\% of all patches in the dataset and belongs to one tail label in RQ1.
We observe that TreeVul achieves an accuracy of 42.8\% on this type, which is significantly lower than the accuracy on the entire dataset (73.1\%).
Moreover, TreeVul wrongly classifies 14.3\% of patches of CWE-661 into CWE-384, CWE-119, CWE-200, and CWE-441, respectively. 
Those erroneous predictions (i.e., CWE-384, CWE-119, CWE-200, and CWE-441) are not relevant to the ground truths CWE-611, indicating that the model struggles to learn the features of this type because of long-tailed distribution.

\begin{tcolorbox} [boxrule=0.8pt,
                top=0.2pt,
                  bottom=0.2pt]
    \textbf{Answer to RQ2}: The experimental results demonstrated that the long-tailed distribution significantly affects the effectiveness of DL models. While these models perform well on head data with frequent data instances, they face difficulties in tail data with infrequent data instances. 
    Across the three tasks, we observed that the performance difference of the models on the head and tail data ranged from 30.0\% to 254.0\%, highlighting the challenge that DL models face in providing accurate recommendations for tail data.
\end{tcolorbox}

\begin{figure}[t]
    \centering
    \begin{subfigure}[b]{0.5\textwidth}
    \centering
      \includegraphics[width=0.85\textwidth]{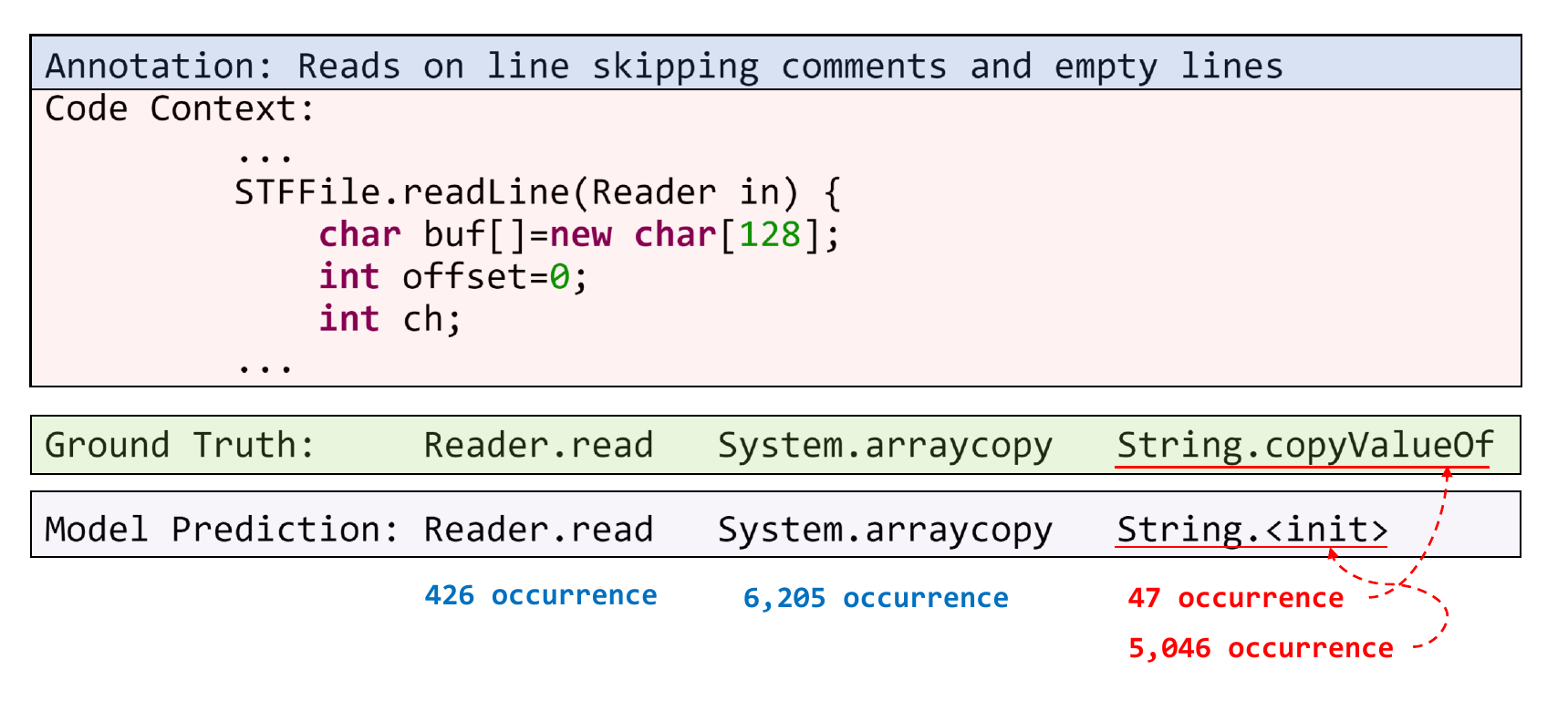}
      \label{fig:qualitative_api}
    \vspace{-0.3cm}
      \caption{API Sequence Recommendation.}
    \end{subfigure}%
    \\
    \begin{subfigure}[b]{0.5\textwidth}
    \centering
      \includegraphics[width=1.0\textwidth]{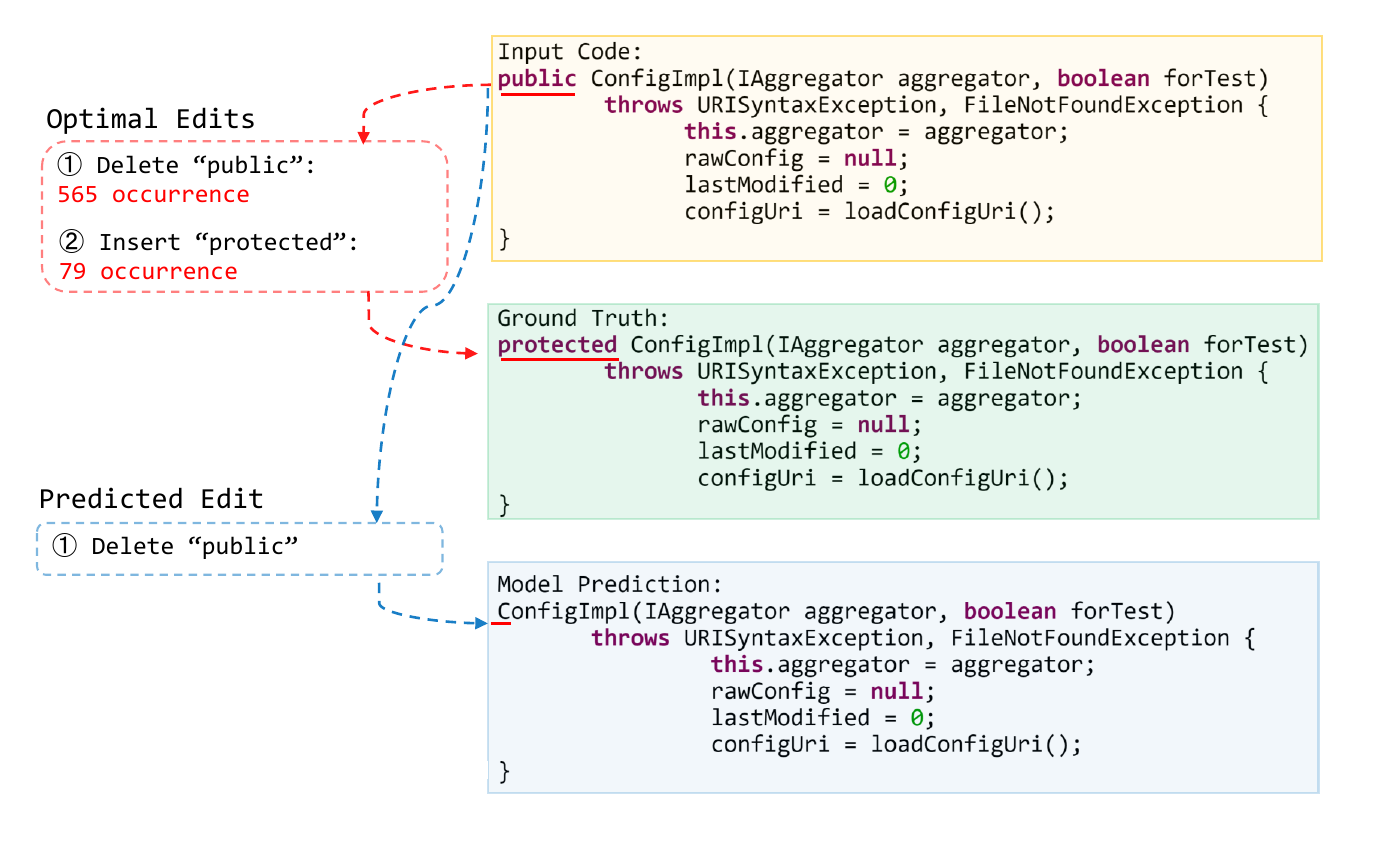}
      \label{fig:qualitative_review}
    \vspace{-0.6cm}
      \caption{Code Revision Recommendation. }
    \end{subfigure}
    \\
    \begin{subfigure}[b]{0.5\textwidth}
      \includegraphics[width=1\textwidth]{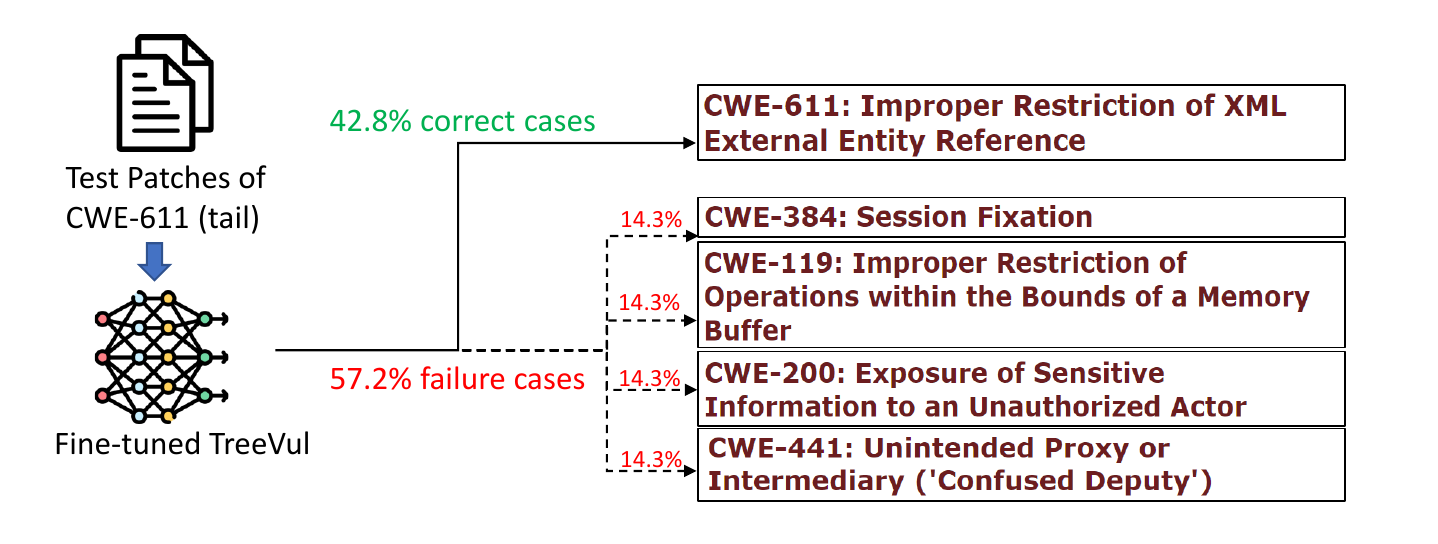}
      \label{fig:qualitative_vul}
    \vspace{-0.3cm}
      \caption{Vulnerability Type Prediction.}
    \end{subfigure}%
    \vspace{-0.1cm}
    \caption{Qualitative analysis of three studied tasks.}
    \label{fig:qualitative}
    \vspace{-0.5cm}
\end{figure}

\subsection{RQ3: Effectiveness of LT Solutions}
\label{sec:result3}

To address the issue of long-tailed distribution, various studies have proposed solutions, e.g., Focal Loss~\cite{focal} and LTR~\cite{lrt_a} techniques, to improve model performance on tail data.
While these solutions were not initially developed for SE datasets, it is still valuable to examine their efficacy on long-tailed SE datasets.

\vspace{0.1cm}
\noindent \textbf{A Specific Challenge in Learning Tail Data.}
Researchers~\cite{yang2022survey_lt} have proposed a theoretical explanation for the difficulty in learning tail data. 
The head classes have a substantially higher number of training samples, which results in the impact of the tail classes being overshadowed by the head classes during model parameter updates.
As a result, the model is trained in a direction that understands heads well, but not tails.

\vspace{0.1cm}
\noindent \textbf{Solutions for Long-tailed Distribution.}
To address the negative impacts of long-tailed distributions, researchers designed various solutions that involve assigning higher weights to tail samples and lower weights to head samples. 
These approaches aim to prevent DL models from being dominated by the head data.
In this research question, we investigate the effectiveness of two different solutions for addressing the long-tailed distribution: the widely used Focal Loss~\cite{focal} and the state-of-the-art solution, namely LTR~\cite{lrt_a}.

\begin{enumerate}[leftmargin=*]
\item \textit{Focal Loss (FL)}~\cite{focal} a classical solution for dealing with long-tailed distributions by introducing an adjustment factor to the standard cross-entropy loss to focus training on difficult samples (i.e., the samples where DL models have low confidence in their predictions). FL increases the loss of difficult samples that a model has low-confidence predictions on. As difficult samples are mostly composed of the tail data for the
long-tailed data, FL improves the learning of tail data~\cite{yang2022survey_lt}. In addition, FL is designed for classification tasks. 
\item \textit{Long-Tailed Recognition via Weight Balancing (LTR)}~\cite{lrt_a} achieved the state-of-the-art results in computer vision tasks. It adopts a two-stage training paradigm: (1) learning features using the standard cross-entropy loss, and (2) further tuning classifiers through class-balanced loss~\cite{cui2019class} with weight decay~\cite{hanson1988comparing} and the MaxNorm constraint~\cite{huang2019deep}, which assigns higher weights to the tail.
\end{enumerate}

Although we investigate one classification task and two generation tasks, the mitigation techniques for long-tailed distribution are for classification tasks. To the best of our knowledge, there is no long-tailed distribution mitigation solution for generation tasks, yet.
Within the context of our study, a significant differentiation between generation tasks and classification tasks is that in generation tasks, the samples have multiple fine-grained labels (APIs/edits), while in classification tasks, the samples have only one label.
On the one hand, we observed that the FL is flexible and does not require a sample with a single label~\cite{focal}. Therefore, we also investigated its effectiveness in generation tasks. On the other hand, LTR uses the class-balanced loss, which assumes that there is only a single label for each sample~\cite{cui2019class}. As a result, we only evaluated its effectiveness in the classification task in this study.

\begin{table}[t]
\centering
\caption{Results of LT solutions in EM scores or accuracy} \label{tab:lt_solution}
\resizebox{0.42\textwidth}{!}{%
    \begin{tabular*}{\linewidth}{l@{\extracolsep{\fill}}*{3}{c}}
    \toprule
    \textsc{Tasks} & \multicolumn{3}{c}{\textsc{Splits}} \\ 
    \toprule
    \textbf{API Seq. Rec.} & Head & Tail & All \\ 
    \midrule
    MulaRec & \textbf{52.0} & \textbf{40.0} & \textbf{46.2} \\ 
    \hspace{0.0em} + FL\cite{focal} & 50.5 (1.5\%\textcolor{dark-red}{$\downarrow$}) & 36.6 (3.4\%\textcolor{dark-red}{$\downarrow$}) & 43.8 (2.4\%\textcolor{dark-red}{$\downarrow$}) \\ 
    \hdashline
    CodeBERT & 46.6 & 30.4 & 38.5 \\  
    \hspace{0.0em} + FL\cite{focal} & 39.5 (7.1\%\textcolor{dark-red}{$\downarrow$}) & 21.2 (9.2\%\textcolor{dark-red}{$\downarrow$}) & 30.4 (8.1\%\textcolor{dark-red}{$\downarrow$}) \\
    \hdashline
    CodeT5 & 51.5 & 38.8 & 45.2 \\
    \hspace{0.0em} + FL\cite{focal} & 47.8 (3.7\%\textcolor{dark-red}{$\downarrow$}) & 36.6 (2.2\%\textcolor{dark-red}{$\downarrow$}) & 42.2 (3.0\%\textcolor{dark-red}{$\downarrow$}) \\
    \toprule
    \textbf{Revision Rec.} & Head & Tail & All \\ 
    \midrule
    T5-Review & 14.1 & 4.1 & 9.0 \\
    \hspace{0.0em} + FL\cite{focal} &  11.8 (2.3\%\textcolor{dark-red}{$\downarrow$}) & 3.2 (0.9\%\textcolor{dark-red}{$\downarrow$}) & 7.4 (1.6\%\textcolor{dark-red}{$\downarrow$}) \\ 
    \hdashline
    CodeBERT & 17.7 & 5.0 & 11.3 \\
    \hspace{0.0em} + FL\cite{focal} & 12.9 (4.8\% \textcolor{dark-red}{$\downarrow$}) & 4.1 (0.9\% \textcolor{dark-red}{$\downarrow$}) & 8.4 (2.9\% \textcolor{dark-red}{$\downarrow$}) \\ 
    \hdashline
    CodeT5 & \textbf{24.6} & \textbf{8.1} & \textbf{16.1} \\ 
    \hspace{0.0em} + FL\cite{focal} & 20.7 (3.9\% \textcolor{dark-red}{$\downarrow$}) & 6.5 (1.6\% \textcolor{dark-red}{$\downarrow$}) & 13.5 (2.6\% \textcolor{dark-red}{$\downarrow$}) \\ 
    \toprule
    \textbf{Vulner. Type} & Head & Tail & All \\ 
    \midrule
    TreeVul & \textbf{87.0} & 60.6 & 73.1 \\
    \hspace{0.0em} + FL\cite{focal} & 82.8 (4.2\% \textcolor{dark-red}{$\downarrow$}) & 59.4 (1.2\% \textcolor{dark-red}{$\downarrow$}) & 70.5 (2.6\% \textcolor{dark-red}{$\downarrow$}) \\ 
    \hspace{0.0em} + LTR\cite{lrt_a} & \textbf{87.0} (0.0\% \textcolor{black}{$-$}) & 61.2 (0.6\% \textcolor{green}{$\uparrow$}) & \textbf{73.4} (0.3\% \textcolor{green}{$\uparrow$}) \\
    \hdashline
    CodeBERT & 82.8 & 59.4 & 70.5 \\
    \hspace{0.0em} + FL\cite{focal} & 81.5 (1.3\% \textcolor{dark-red}{$\downarrow$}) & \textbf{61.7} (2.3\% \textcolor{green}{$\uparrow$}) & 71.1 (0.6\% \textcolor{green}{$\uparrow$}) \\ 
    \hspace{0.0em} + LTR\cite{lrt_a} & 82.8 (0.0\% \textcolor{black}{$-$}) & 60.3 (0.9\% \textcolor{green}{$\uparrow$}) & 70.9 (0.4\% \textcolor{green}{$\uparrow$}) \\
    \hdashline
    CodeT5 & 81.6 & 50.7 & 65.3 \\ 
    \hspace{0.0em} + FL\cite{focal} & 80.3 (1.3\% \textcolor{dark-red}{$\downarrow$}) & 53.4 (2.7\% \textcolor{green}{$\uparrow$}) & 65.9 (0.6\% \textcolor{green}{$\uparrow$}) \\ 
    \hspace{0.0em} + LTR\cite{lrt_a} & 80.3 (1.3\% \textcolor{dark-red}{$\downarrow$}) & 54.5 (3.8\% \textcolor{green}{$\uparrow$}) & 66.7 (1.4\% \textcolor{green}{$\uparrow$}) \\

    \bottomrule
    \end{tabular*}
    }
    \label{tab:lt_solution}
    \begin{flushleft}
    \begin{footnotesize}
    {*API Seq. Rec.: API sequence recommendation, Revision Rec.: code revision recommendation, and Vulner. Type: vulnerability type prediction.}
    \end{footnotesize}
\end{flushleft}
\end{table}

\vspace{0.1cm}
\noindent \textbf{Experimental Results}.
The aforementioned mitigation solutions are strategies that re-weight the standard cross-entropy loss based on the label frequencies. Therefore, we apply them to the studied DL-based SE approaches.
Table~\ref{tab:lt_solution} displays the performance of the SE models after integrating the potential solutions. 
We highlight the best performance in each split (head/tail/all) for each task by using boldface numbers. The results indicate that FL consistently diminishes the effectiveness of SE models in generation tasks, i.e., API sequence recommendation and code revision recommendation, regardless of the split. 
Specifically, the integration of FL results in a 1.5--9.2\% and 0.9--4.8\% absolute percentage reduction in EM for API sequence recommendation and code revision recommendation, respectively.

Regarding the vulnerability type prediction, we found that FL and LTR resulted in enhanced accuracy on the tail for most cases. Specifically, FL yielded a 2.3\% and 2.7\% absolute percentage increase in tail accuracy for CodeBERT and CodeT5, respectively, while it caused a 1.2\% decrease in tail accuracy for TreeVul. 
In contrast, LTR consistently improved the performance on the tail for all models by an absolute percentage of 0.6--3.8\%. Interestingly, we observed that FL sacrifices the performance on the head to achieve a potential improvement on the tail while LTR can maintain the head performance for CodeBERT and TreeVul.
In summary, the studied long-tailed solutions could improve the tail performance of the vulnerability type prediction by 0.6--3.8\%. 
Nevertheless, these solutions are not guaranteed to produce consistent enhancements and only have a marginal effect on the overall results of the complete test set, ranging from 0.3\% to 1.4\%.

\vspace{0.1cm}
\noindent \textbf{Properties of SE datasets}.
Based on the experimental results, we found that the mitigation solutions are not effective enough for software engineering datasets. 
It possibly stems from the fact that these solutions treat each label as completely distinct, ignoring the relationships among them. 
In computer vision datasets, labels such as ``plane'' and ``cart'' in image classification are mostly independent and it is not common to analyze the label relationships~\cite{imagenet-lt}.
In contrast, software engineering datasets contain rich information regarding relationships among different labels. For instance, a specific Common Weakness Enumeration (CWE) type can have several sibling CWE types that share the same parent type in the CWE type tree and describe similar vulnerabilities~\cite{treevul}. 
An API usually is used as a part of combinations with other APIs~\cite{mularec}. Token-level edits may occur simultaneously as shown in Figure~\ref{fig:qualitative}(b). 
Moreover, the relationship information among different labels in SE datasets can be obtained from historical data (e.g., API usages or code changes) or expert knowledge-based systems maintained by the community (e.g., the CWE System).

\begin{tcolorbox} [boxrule=0.8pt,
                top=0.2pt,
                  bottom=0.2pt]
    \textbf{Answer to RQ3}: 
    The widely-used solutions for addressing long-tailed distributions have proven to be relatively ineffective in SE datasets, resulting in only marginal improvements ranging from 0.3\% to 1.4\% across the complete datasets. 
    This challenge highlights the difficulty of effectively learning the tail part of SE datasets. 
    Thus, we encourage future research to explore more potential solutions to this issue.

\end{tcolorbox}
\subsection{RQ4: Tail Data Detection}
\label{sec:rq4}

Identifying tail data is potentially useful in warning users about the reliability of generated predictions of learning-based SE tools. 
Therefore, this research question seeks to explore the effectiveness of identifying tail data solely based on input data.

\begin{table}[b]
\centering
\caption{Results of identifying the tail}
\resizebox{0.5\textwidth}{!}{%
\begin{tabular}{l|c|c|c}
\hline
\multicolumn{1}{l|}{\textbf{Datasets}} & \textbf{API Seq. Rec.} & \textbf{Revision Rec.}  & \textbf{Vulner. Type Pred.} \\ \hline
\textbf{Accuracy}                  & 84.4                                                              & 66.7                                                             & 81.7                                                                                                               \\ \hline
\textbf{F1}                        & 84.0                                                              & 66.4                                                             & 81.5                                                                         \\ \hline
\end{tabular}
}
\label{tab:rq4}
\end{table}

\vspace{0.1cm}
\noindent \textbf{Experimental Setup}.
We conduct a further investigation by fine-tuning the CodeBERT model to perform binary classification, i.e., predicting the probability of input data belonging to either the tail or the head. Specifically, we consider the tail as the positive class and the head as the negative class. We train CodeBERT to minimize cross-entropy loss.

\vspace{0.1cm}
\noindent \textbf{Results}.
As shown in Table~\ref{tab:rq4}, the fine-tuned CodeBERT demonstrated an ability to accurately identify the tail data in the datasets of API sequence recommendation and vulnerability type prediction, achieving accuracy scores of 84.4\% and 81.7\%, respectively. 
However, its performance on the code revision recommendation dataset was substantially lower with an accuracy of 66.7\%.

\vspace{0.1cm}
\noindent \textbf{Potential Application}.
Recent research has highlighted that incorrect prediction outcomes (i.e., false positives) can result in developers' discontent and reduced trust in automated SE approaches~\cite{johnson2013don,kochhar2016practitioners,le2018overfitting}. 
We are aware that SE models show particularly lower effectiveness on tail data as addressed in RQ2 and RQ3. 
One potential utility of the tail data detection tool is to either explicitly advise users to disregard predictions on the tail data or compel automated SE tools to output a response like ``\textit{I am uncertain about the answer due to lack of relevant knowledge}.'' 
Such an application might enhance user satisfaction by aiding in the identification and filtering out of prediction results regarding tail data that are more prone to be inaccurate.

\begin{tcolorbox} [boxrule=0.8pt,
                top=0.2pt,
                  bottom=0.2pt]
    \textbf{Answer to RQ4}: Experimental results demonstrated accurate identification (84.4\% and 81.7\% in accuracy) of the tail in API sequence recommendation and vulnerability type prediction. 
    However, our approach achieves a lower accuracy of 66.7\% in code revision recommendations. 
    This suggests that the difficulty of identifying the tail varies depending on a task and the associated dataset.
\end{tcolorbox}

\section{Discussion}
\label{sec:discussion}

\subsection{How Do Traditional Machine Learning Models Perform?}

In addition to the previously studied DL models, there exists another category of widely used methods known as traditional Machine Learning (ML) models. In our evaluation, we take popular traditional machine learning models such as Logistic Regression (LR)~\cite{zhu1997algorithm}, Decision Tree (DT)~\cite{hastie2009elements}, Support Vector Machines (SVM)~\cite{crammer2001algorithmic}, and Random Forest (RF)~\cite{breiman2001random} into account.
However, it is important to note that traditional ML models are solely designed for classification tasks. Therefore, we solely perform experiments on the vulnerability type prediction task. 
Specifically, we adopt the widely used Bag-of-Word (also called Bag-of-Token) embedding~\cite{ko2012study} to turn the input data into vectors and feed the vectors into the traditional ML models.
Table~\ref{tab:traditional_ml} illustrates the performances of traditional ML models (numbers in grey) on the head, tail, and whole data in accuracy. 
We also included the performance of the state-of-the-art approach, TreeVul, in this table. The results show that similar to DL, traditional ML models perform better on the head data compared to the tail data.

\begin{table}[t]
\caption{Performance of traditional machine learning models on vulnerability type prediction}
\centering
\begin{tabular}{l|
>{\columncolor[HTML]{EFEFEF}}r |
>{\columncolor[HTML]{EFEFEF}}r |
>{\columncolor[HTML]{EFEFEF}}r |
>{\columncolor[HTML]{EFEFEF}}r |r}
\hline
\textbf{Acc.} & \multicolumn{1}{c|}{\cellcolor[HTML]{EFEFEF}\textbf{LR}} & \multicolumn{1}{c|}{\cellcolor[HTML]{EFEFEF}\textbf{SVM}} & \multicolumn{1}{c|}{\cellcolor[HTML]{EFEFEF}\textbf{DT}} & \multicolumn{1}{c|}{\cellcolor[HTML]{EFEFEF}\textbf{RF}} & \multicolumn{1}{c}{\textbf{TreeVul}} \\ \hline
\textbf{Head} & 60.0                                                     & 53.8                                                      & 51.6                                                     & 63.1                                                     & 87.0                                 \\ \hline
\textbf{Tail} & 40.7                                                     & 41.6                                                      & 38.2                                                     & 43.4                                                     & 60.6                                 \\ \hline
\textbf{All}  & 50.2                                                     & 47.6                                                      & 44.8                                                     & 53.1                                                     & 73.1                                 \\ \hline
\end{tabular}
\label{tab:traditional_ml}
\end{table}

\subsection{Implications}
\label{sec:implication}

\textbf{Researchers need to be cautious about using average results.}
In the field of SE, it is common practice to report average results, such as average accuracy, on all samples in the test set. 
However, this practice inadvertently conceals the model's shortcomings on the tail, as it performs exceptionally well on the head data but poorly on the tail data. 
For example, in vulnerability type prediction, where there are 117 categories, TreeVul achieves an accuracy of 60.6\% on the majority of the types (107 tail types), but due to the high accuracy of the 6 head types (87.0\%), the average result is 73.1\%. 
However, TreeVul cannot reach an accuracy of 73.1\% on most of the CWE types (107 tail types) in this task. 
To gain a more comprehensive understanding of both the strengths and limitations of an approach, we recommend that researchers report both the average results and the results on heads and tails separately if long-tailed distributions exist in their datasets.

\textbf{A customized method to effectively learn the tail in software engineering data is needed. Our future work will take the rich label relationships into account. } 
Our study sheds light on the ineffectiveness of adopting computer vision solutions for long-tailed distribution problems in SE datasets. 
One possible reason is that computer vision solutions do not account for the rich label relationships in SE data. These relationships can be valuable resources for learning the features of tail labels. 
A potential approach is to group similar tail labels into clusters, referred to as abstracted classes, which could provide more data samples and improve the learning of effective features for the abstracted class that consists of tail labels.

\subsection{Threats to Validity} 
\label{sec:threats}
Our findings are subjected to the studied models and datasets. Therefore, it may not be generalizable to all software engineering tasks. To mitigate this limitation, we selected three distinct tasks to cover two representative categories of software engineering tasks, i.e., classification and generation tasks.
The findings of our study for RQ2 are based on the even division of test data into head and tail data, which may not be the optimal approach for uncovering insights into long-tailed code distributions. To address this limitation, we also split the test data into 10 small groups. This splitting helped to mitigate the potential bias and confirmed the consistency of our findings.
Our study primarily focused on DL-based approaches. The generalizability of our findings to traditional machine learning models, such as Logistic Regression, is uncertain.
To address this limitation, we conducted experiments using four traditional machine learning models. 
Finally, we would like to note that we made our replication package~\footnote{\url{https://github.com/soarsmu/LT4Code}} publicly available for future studies to repeat our experiments and extend our work.

\section{Conclusion and Future Work}
\label{sec:conclusion}
We presented an empirical study on the long-tailed distribution of SE data. Specifically, we investigated this phenomenon on three distinct SE downstream tasks, namely API sequence recommendation, code revision recommendation, and vulnerability type prediction. 
Our study revealed that the long-tailed distribution has a significant impact on the performance of deep learning-based SE approaches. Specifically, deep learning models performed 30.0\% to 254.0\% worse on tail data compared to head data.
In this study, we explored the effectiveness of solutions designed to mitigate the negative impact of the long-tailed data distribution in images. However, we found that these solutions are not effective enough in addressing long-tailed distribution in SE data, and we call for future work to address this issue.

Moving forward, we plan to extend our investigation to more SE tasks and design a novel approach to addressing long-tailed distribution in code by considering the inner relationships among labels. 
Moreover, we are also interested in utilizing recent popular generative AI models like ChatGPT to generate samples for the tail classes, thereby mitigating the issue of long-tailed distributions in some SE datasets.

\vspace{0.1cm}
\noindent{\bf Acknowledgement.} This research / project is supported by the National Research Foundation, Singapore, under its Industry Alignment Fund – Pre-positioning (IAF-PP) Funding Initiative. Any opinions, findings and conclusions or recommendations expressed in this material are those of the author(s) and do not reflect the views of National Research Foundation, Singapore.

\bibliographystyle{IEEEtran}
\bibliography{sample-base}

\end{document}